\documentclass[aps,prl,preprint,onecolumn,superscriptaddress,nobibnotes,amsmath,amssymb]{revtex4-2}

\usepackage{graphicx}
\usepackage{color}
\usepackage[explicit]{titlesec}
\usepackage{hyperref}

\begin{document} 
	
	\title{Heat-Mode Excitation in a Proximity Superconductor}
\author{A.O.~Denisov}
\affiliation{Osipyan Institute of Solid State Physics, Russian Academy of Sciences, 142432 Chernogolovka, Russia}
\affiliation{Department of Physics, Princeton University, Princeton, New Jersey 08544, USA}
\author{A.V.~Bubis} 
\affiliation{Skolkovo Institute of Science and Technology, Nobel street 3, 121205 Moscow, Russian Federation}
\affiliation{Osipyan Institute of Solid State Physics, Russian Academy of Sciences, 142432 Chernogolovka, Russia}
\author{S.U.~Piatrusha}
\affiliation{Osipyan Institute of Solid State Physics, Russian Academy of Sciences, 142432 Chernogolovka, Russia}
\author{N.A.~Titova}
\affiliation{Moscow Pedagogical State University, 29 Malaya Pirogovskaya St, 119435 Moscow, Russia}
\author{A.G.~Nasibulin}
\affiliation{Skolkovo Institute of Science and Technology, Nobel street 3, 121205 Moscow, Russian Federation}
\author{J.~Becker}
\affiliation{Walter Schottky Institut, Physik Department, and Center for Nanotechnology and Nanomaterials, Technische Universit\"{a}t M\"{u}nchen, Am Coulombwall 4, Garching 85748, Germany}
\author{J.~Treu}
\affiliation{Walter Schottky Institut, Physik Department, and Center for Nanotechnology and Nanomaterials, Technische Universit\"{a}t M\"{u}nchen, Am Coulombwall 4, Garching 85748, Germany}
\author{D.~Ruhstorfer}
\affiliation{Walter Schottky Institut, Physik Department, and Center for Nanotechnology and Nanomaterials, Technische Universit\"{a}t M\"{u}nchen, Am Coulombwall 4, Garching 85748, Germany}
\author{G.~Koblm\"{u}ller}
\affiliation{Walter Schottky Institut, Physik Department, and Center for Nanotechnology and Nanomaterials, Technische Universit\"{a}t M\"{u}nchen, Am Coulombwall 4, Garching 85748, Germany}
\author{E.S.~Tikhonov}
\affiliation{Osipyan Institute of Solid State Physics, Russian Academy of Sciences, 142432 Chernogolovka, Russia}
\author{V.S.~Khrapai}
\affiliation{Osipyan Institute of Solid State Physics, Russian Academy of Sciences, 142432 Chernogolovka, Russia}
\affiliation{National Research University Higher School of Economics, 20 Myasnitskaya Street, 101000 Moscow, Russia}

	\begin{abstract}
	Mesoscopic superconductivity deals with various quasiparticle excitation modes, only one of them---the charge-mode---being directly accessible for conductance measurements due to the imbalance in populations of quasi-electron and quasihole excitation branches. Other modes carrying heat or even spin, valley etc. currents  populate the branches equally and are charge-neutral, which makes them much harder to control. This noticeable gap in the experimental studies of mesoscopic non-equilibrium superconductivity can be filled by going beyond the conventional DC transport measurements and exploiting spontaneous current fluctuations. Here, we perform such an experiment and investigate the transport of heat in an open hybrid device based on a superconductor proximitized InAs nanowire. Using shot noise measurements, we investigate sub-gap Andreev heat guiding along the superconducting interface and fully characterize it in terms of the thermal conductance on the order of $G_\mathrm{th}\sim e^2/h$, tunable by a back gate voltage. Understanding of the heat-mode also uncovers its implicit signatures in the non-local charge transport. Our experiments open a direct pathway to probe generic charge-neutral excitations in superconducting hybrids.
\end{abstract}

\maketitle

\section{Introduction}

Conversion of a quasiparticle current to the collective motion of a Cooper pair condensate at the interface of a normal metal and superconductor is known as Andreev reflection (AR)~\cite{Andreev}. For~quasiparticle energies {($\varepsilon$)} below the superconducting gap  {($\Delta$)} (sub-gap quasiparticles, {$|\varepsilon|<\Delta$}), AR is fully responsible for the charge transport across the interface. Conservation of both the number of sub-gap quasiparticles and their excitation energy on the normal side manifests AR as a fundamental example of charge--heat separation in the electronic system. Out of thermal equilibrium, the~spatial gradient of a charge-neutral quasiparticle distribution conveys the heat flux~\cite{RevModPhys.78.217}, which does not penetrate the superconductor and propagates along its boundary with a normal conductor. In~this way, ARs mediate the heat conduction via vortex core in s-type superconductors~\cite{Caroli_1964} and via neutral modes in graphene~\cite{PhysRevB.75.045417}.

The retro-character of the AR, that is, the propagation of a reflected hole via the time-reversed trajectory of an incident electron, results in a suppression of the heat conduction in the ballistic limit. This obstacle may be overcome by imposing the chirality of the charge carriers in a magnetic field~\cite{Lee_2017,Zhao_2020,Kurilovich_2022}, similar to quantum Hall-based experiments~\cite{Banerjee2017}, or~by going in the regime of specular AR near charge-neutrality point in graphene~\cite{Specular_AR_2006}. In~the diffusive limit, counter-intuitively, the~heat transport is restored, since moderate disorder scattering effectively increases the number of the conducting modes~\cite{Kopnin_2004}. In~addition, the~disorder scattering promotes the relaxation of a charge-mode component into pure heat-mode, by~mixing the quasi-electron and quasihole branches via AR. For~such a relaxation to occur, a superconducting gap has to vary either in momentum space, as~in anisotropic bulk superconductors~\cite{tinkham2004introduction}, or~in real space~\cite{Artemenko_UFN_1979}, as~in proximity structures, including in the present experiment. All of this makes the geometry of the Andreev wire~\cite{Kopnin_2004}---a diffusive normal core proximitized by a wrapped around superconductor---preferable for a sub-gap heat transport~experiment.

In this work, we challenge a thermal conductance ($G_\mathrm{th}$) measurement in an open three-terminal hybrid device based on a diffusive InAs nanowire (NW) proximitized by a superconducting contact, see the image of one of our samples in Figure~\ref{fig1}a. Conceptually similar devices were investigated in the context of Cooper-pair splitters~\cite{Hofstetter_2009,Herrmann2010,DasDas2012} and, more recently, Majorana physics~\cite{PhysRevLett.105.077001,PhysRevLett.105.177002,Albrecht2016,Deng1557,yu2020nonmajorana,Wang_2021} with the emphasis on the electrical conductance. The~central part of the device represents a few 100~nm long  Andreev wires with a partial superconducting wrap, which removes complications arising from the Little--Parks effect~\cite{Vaitiek_nas_2020,Kopasov_PhysRevB.101.054515}. In~a previous work with the same devices~\cite{Denisov_SST2021}, we have demonstrated a charge neutrality of a non-local quasiparticle response, which is direct evidence of the heat-mode excitation regime. Here, we focus on a comparison of local and non-local noise signals, evaluation of thermal conductance and the origin of transport signals in this regime. Our experiments offer a so-far missing experimental tool in the field of non-equilibrium mesoscopic superconductivity~\cite{Keizer2006,Huebler_2010,Vercruyssen_2012,Golikova_2014,RevModPhys.90.041001,kuzmanovi2020evidence} and enable the control of generic charge-neutral excitations in superconducting~hybrids.

\section{Results: Devices and Transport~Response} 

\begin{figure}[h]
 		\includegraphics[width=13.5 cm]{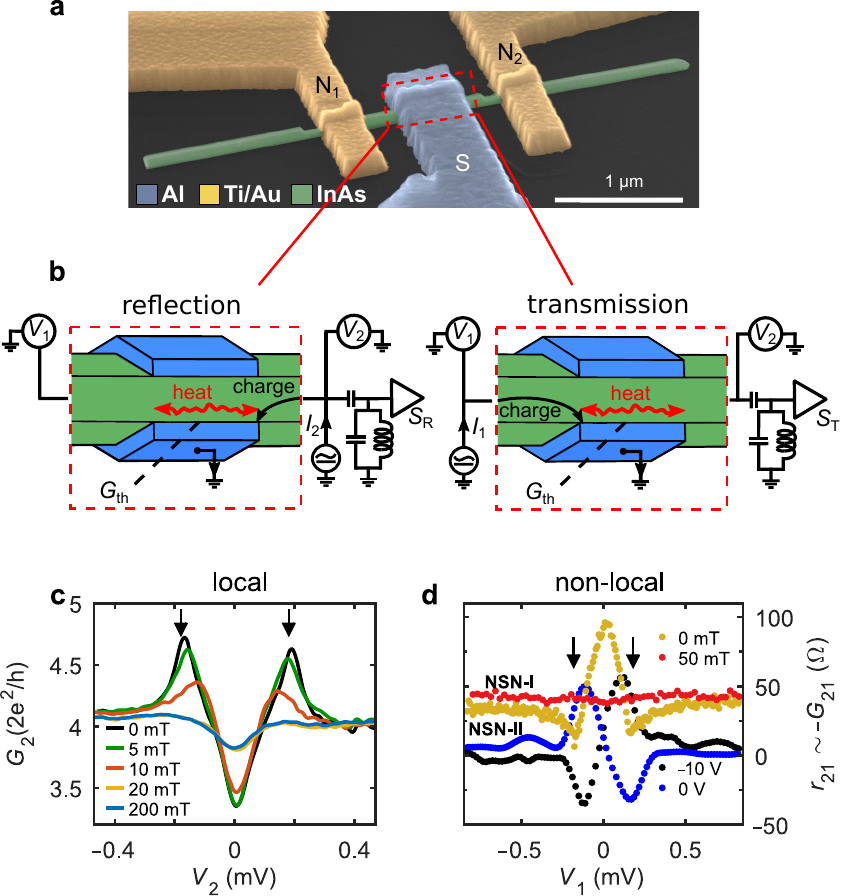}
 	\caption{{Outline and charge transport data.} (\textbf{a})~Scanning electron microscope image of the typical device (false color). InAs NW is equipped with two N terminals (Ti/Au) on the sides and one S-terminal (Al) in the middle. (\textbf{b})~Separation of charge and heat currents at the InAs/Al interface {and two noise measurement configurations}. \mbox{The three-terminal} {device layout} allows studying thermal conductance $G_{\mathrm{th}}$ of the proximitized NW region by measuring shot noise {in the transmission configuration}. Note that in {the present experiment, only terminal N2} is connected to the low temperature amplifier, so that switching between the reflection noise $S_\mathrm{R}$ and transmission noise $S_\mathrm{T}$ is achieved by interchanging the biased and floating N-terminals, see the Supplemental Materials for the wiring scheme. (\textbf{c})~Local differential conductance of NS junction in device NSN-II measured at $T=50~\mathrm{mK}$ in different magnetic fields. (\textbf{d})~Non-local differential resistance $r_{21}\equiv dV_{\mathrm{2}}/dI_{\mathrm{1}}$ for two devices plotted at different $B$ and $V_{\mathrm{g}}$.}
 	\label{fig1}
 \end{figure}

{The outline of our experiment is depicted in Figure~\ref{fig1}b. A~semiconducting InAs nanowire is equipped with a superconducting (S) terminal, made of Al, in~the middle and two normal metal (N) terminals, made of Ti/Au bilayer, on~the sides. {Below, we focus on the data from two devices. In~the device NSN-I (NSN-II), the~length of the NW underneath the superconductor is 200~nm (300~nm) and the NW segments between the S-terminal and the N-terminals are 350~nm (300~nm) long.} In essence, this device {layout} represents two back-to-back normal metal--NW--superconductor (NS) junctions sharing the same S-terminal. Note the absence of the quantum dots~\cite{Hofstetter_2009,DasDas2012,Deng1557} or tunnel barriers~\cite{Albrecht2016} adjacent to the S-terminal, which enables better coupling of the sub-gap states to the normal conducting regions. Throughout the experiment, the S-terminal is grounded, terminal N1 is biased and terminal N2 is floating (or vice~versa). Note that grounding of the S-terminal protects the Al from non-equilibrium superconductivity effects~\cite{Keizer2006,Bubis_2021}. The~S-terminal serves as a nearly perfect sink for the charge current. At~energies below the superconducting gap $\Delta\approx180~\mu$eV of Al, the S-terminal cannot absorb quasiparticles~\cite{Andreev} and their non-equilibrium population can relax only via diffusion to the N terminals~\cite{PhysRevB.63.081301}, manifesting charge--heat separation. This charge-neutral diffusion flux, which is referred to as the heat flux below, is shown by curly arrows in Figure~\ref{fig1}b. One part of the heat flux relaxes via {the biased} terminal, similar to the usual two-terminal configurations~\cite{PhysRevLett.84.3398,Jehl2000}. The~other part bypasses the S-terminal and relaxes via floating terminal. As~we will demonstrate below, this heat flux can be detected by means of shot noise~thermometry.}

For charge--heat separation via AR, the quality of the InAs/Al interface is important, which we verify in transport measurements. In~Figure~\ref{fig1}c, we show the local differential conductance $G_2$ of the biased junction N2-S in device NSN-II as a function of voltage $V_2$ at a temperature $T=50~$mK. Without~the magnetic field ($B$), $G_2$ exhibits two well-defined maxima at finite $V_2$ that diminish with increasing the $B$-field directed perpendicular to the substrate and vanish in $B\approx20~$mT simultaneously with the transition of the Al to {the} normal state. The~maxima occur around gap edges $V_2=\pm\Delta/|e|$, where $e$ is the elementary charge, and~the corresponding increase of $G_2$ above the normal state value reaches about 15\%. This re-entrant conductance behavior is a property of diffusive NS junctions with a highly transparent interface~\cite{Courtois1999}. Around zero bias in $B=0$, we generally observed a small reduction of $G_2$ by about 10\% in all back gate voltage ($V_\mathrm{g}$) range studied. This guarantees that possible residual reflectivity has a minor effect and ARs dominate over normal interface scattering in our~devices.

In Figure~\ref{fig1}d, we plot non-local differential resistance $r_\mathrm{21}=dV_2/dI_1$, where $V_2$ is the voltage on terminal N2, as~a function of $V_1$. In~the normal state $r_\mathrm{21}$ is featureless and consists of the interface resistance along with a few-Ohm contribution of the Al lead, see the trace in $B=50$~mT in device NSN-I with $r_\mathrm{21}\approx 40~\Omega$. By~contrast, in~$B=0$ strong gap-related features develop and $r_\mathrm{21}$ demonstrates local maximum and minima at the gap edges, see vertical arrows. Note that $B=0$ behavior is non-universal and depending on $V_\mathrm{g}$, we have also observed bias asymmetry and sign reversal of the $r_\mathrm{21}$, see two lower datasets for the device NSN-II. {These features are related to the energy dependence of the sub-gap conductance and}  have {a} thermoelectric-like origin~\cite{Eom_1998}, as will be discussed below. Overall, $r_\mathrm{21}$ being small compared to the individual resistances of the NS junctions signals that the current transfer length $l_\mathrm{T}$ is small compared with the width of the S-terminal. We estimate $l_\mathrm{T}\leq100~$nm close to the superconducting coherence length in Al, which sets the lowest possible bound for the $l_\mathrm{T}$, see Supplemental Materials for the details. $r_\mathrm{21}$ can be expressed via a non-diagonal element of the conductance matrix~\cite{PhysRevB.97.045421} as $r_\mathrm{21}\approx-G_{21}/G_{2}G_{1}$, where $G_{i}\approx G_{ii}$ ($i=1,2$) are the two-terminal conductances of the NS junctions. $G_{21}\sim 10^{-2} G_{i}$ is a direct consequence of a charge-neutrality of the non-local response in our devices~\cite{Denisov_SST2021} and proves nearly perfect efficiency of the S contact as charge current sink. {The actual sign of the non-local conductance $G_{21}$ can be both negative and positive, as~determined by a competition of normal and Andreev transmission processes. Corresponding non-local transmission probabilities are commonly denoted by $T_{21}^{ee}$ and $T_{21}^{he}$, respectively~\cite{Anantram_1996}. In~the present experiment,  at~zero bias, we observe a small negative conductance $G_{21}<0$, implying that $\Sigma T_{21}^{ee}>\Sigma T_{21}^{he}$, where sum is performed over the eigenchannels.}

\section{Results: Shot Noise~Response} 

Next, we probe the non-equilibrium electronic populations in both NS junctions using shot noise current fluctuations picked-up {in the reflection and transmission configurations sketched in Figure~\ref{fig1}b}. {This measurement is performed using a schematics based on a resonant tank circuit and a home-made low-temperature amplifier. The~measurement layout and the calibration procedure are detailed in the Supplemental Materials.} Figure~\ref{fig2}a demonstrates the noise spectral density measured in terminal N2 as a function of $I_2$ at two gate voltages. This configuration, referred to as the reflection configuration, is reminiscent of the usual AR noise in two-terminal devices~\cite{PhysRevLett.84.3398,Jehl2000}, and~the measured noise is denoted as $S_\mathrm{R}$.
{Experimentally, $S_\mathrm{R}$ represents the spectral density of the auto-correlation noise of current $I_2$ under the bias applied to the terminal N2, while the terminal N1 is maintained DC  floating, that is, $S_\mathrm{R}\equiv S_{22}(I_1=0,I_2)$. The~corresponding experimental layout is depicted in the left sketch of Figure~\ref{fig1}b.} For comparison, a~similar measurement in a reference NS device is shown in Figure~\ref{fig2}c. In~both devices, the results are qualitatively similar, that is, the $S_\mathrm{R}$ scales linearly with current and exhibits clear kinks at the gap edges (marked by the arrows). Above~the kinks, the~diminished slope is the same and it corresponds closely to the universal Fano factor $F\equiv1/3$ in a diffusive conductor with normal leads~\cite{PhysRevLett.76.3806,PhysRevB.59.2871} $\delta S_\mathrm{R}/2e\delta I\approx F$, as~shown by the dashed lines with a marker ``{\it e}''. This familiar behavior~\cite{DasDas2012} verifies elastic diffusive transport in InAs NWs~\cite{Tikhonov2016} even at energies well above $\Delta$ and ensures quasiparticle relaxation solely by diffusion in contacts. {In particular, this observation establishes a solid correspondence between the applied bias voltage and the quasiparticle excitation energy in the present experiment. Namely, a~small bias window of $[V;V+dV])$ corresponds to a creation of electron-like and hole-like quasiparticles with the excess energy of $|\varepsilon|=|eV|$.} {At} sub-gap biases {($|V|<\Delta/|e|$)}, we observe an important difference being a result of joining an extra N-terminal. While in the NS device the slope expectedly doubles~\cite{PhysRevLett.84.3398,DasDas2012}, see the dotted lines in Figure~\ref{fig2}c with the effective charge $e^*\approx 2e$ denoted by ``{\it 2e}'', in~the NSN device, it increases much more weakly and corresponds to $e^*\approx 1.6e$ assuming the same $F$. Unlike in SNS junctions~\cite{Ronen_2016}, a~fractional value of $e^*$ here is not related to a quasiparticle charge in the superconductor, but~reflects an unusual boundary condition for the heat flux underneath the S-terminal, see Ref.~\cite{Bubis_2021} and Supplemental Materials for the details. While the doubled $e^*$ is a direct consequence of the full reflection of heat flux at the S-terminal~\cite{PhysRevB.63.081301}, its intermediate value means that the missing heat flux in the NSN device is transmitted towards the nearby floating N-terminal. Similar behavior was previously observed in topological insulators~\cite{PhysRevLett.117.147001}, however, in~the present experiment, the transmitted heat flux is directly measurable, as we show~below.

\begin{figure}[h]
		\includegraphics[width=13.5 cm]{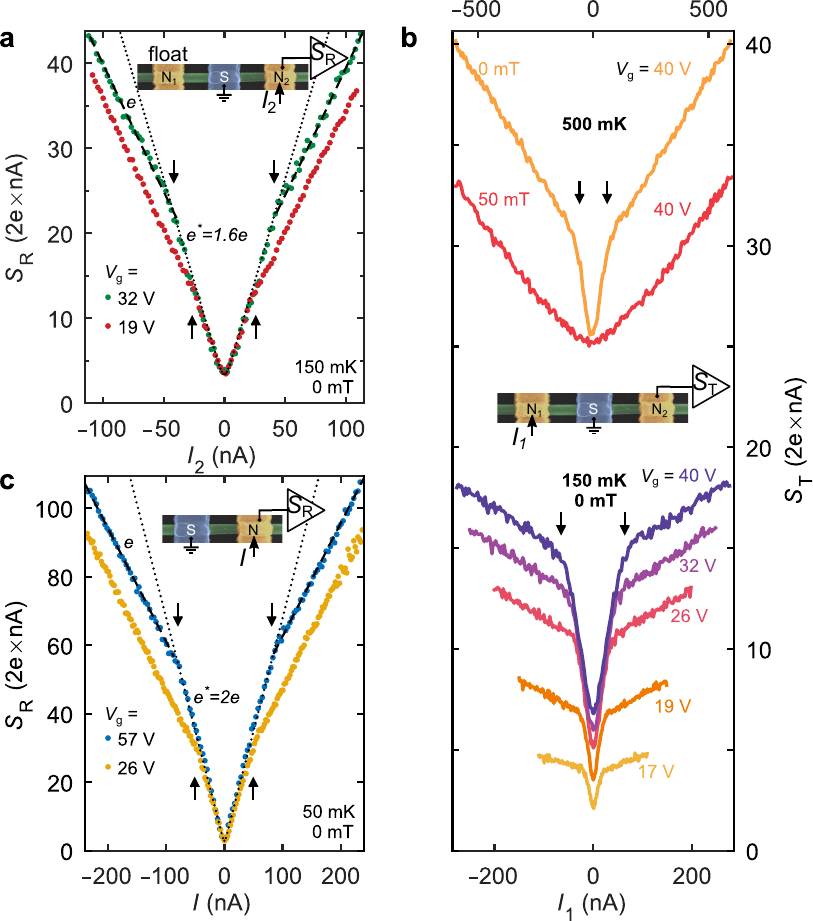}
	\caption{{Reflected and transmitted shot noise}. (\textbf{a})~Reflection noise configuration in device NSN-I. Noise spectral density of the biased NS junction as a function of current at two values of $V_\mathrm{g}$. Dotted line is the fit with $F=0.30$ and charge $e^*=1.6e$; dashed line slope corresponds to $F=0.30$ and charge equal to $e$. Green symbols are shifted vertically by $9\times~10^{-28}~\mathrm{A^2/Hz}$ 
to coincide with red ones at zero bias. (\textbf{b})~Transmission noise configuration in device NSN-I. Noise spectral density of the floating NS junction as a function of current at different $B$, $T$ and $V_\mathrm{g}$ (see legend). (\textbf{c})~Reflected shot noise in the reference two-terminal NS device as a function of current at two values of $V_\mathrm{g}$. Dotted line is the fit with $F=~0.33,~e^*=~2e$; dashed line slope corresponds to $F=0.33$ and charge equal to $e$. }
	\label{fig2}
\end{figure}

In Figure~\ref{fig2}b, we plot the current dependencies of the shot noise measured in transmission configuration, $S_\mathrm{T}$, that is, the noise at the floating terminal N2. {In this configuration, we measure the auto-correlation noise at the DC floating terminal N2 under a finite bias current $I_1$, that is, $S_\mathrm{T}\equiv S_{22}(I_1,I_2=0)$. The~corresponding experimental layout is depicted in the right sketch of Figure~\ref{fig1}b.} Within all investigated $V_\mathrm{g}$ range, $S_\mathrm{T}$ steeply increases at small currents followed by pronounced kinks at the gap edges, see the arrows for some of the traces, and~keeps increasing much more weakly above the kinks. This behavior of $S_\mathrm{T}$ is explained as follows. Sub-gap quasiparticles diffusing along the superconductor, and~experiencing a few ARs on the way, guide the heat flux via proximitized InAs. Above-gap quasiparticles, however, mostly leave via the S-terminal and their contribution to the transmitted heat signal is minimal. This qualitative picture is proved in the following crosscheck experiment. In~the upper part of Figure~\ref{fig2}b, the~$S_\mathrm{T}$ signals are compared in $B=0$ and $B=50$~mT with the Al in superconducting and normal states, respectively. In~the normal state, $S_\mathrm{T}$ grows weakly at increasing $I_1$ without any kinks. Moreover above-gap signal in $B=0$ {roughly} reproduces this trend up to a vertical shift at high $I_1$. We conclude that this effect is {mainly} caused by residual normal interface scattering, see also Ref.~\cite{Denisov_SST2021}. {Importantly,} for sub-gap energies, $S_T\sim S_R$, cf. Figure~\ref{fig2}a, whereas non-local charge transport resulted in $|G_{21}|\ll G_1,G_2$. This difference emphasizes the fact that non-equilibrium populations of quasiholes and quasi-electrons are balanced in the proximity region and transmitted noise directly probes the heat-mode excitation. {Figure~\ref{fig2}b, therefore, demonstrates our main result that at sub-gap energies the proximitized InAs NW supports guiding of heat underneath the S-terminal by virtue of AR processes.}

\begin{figure}[h]
		\includegraphics[width=13.5 cm]{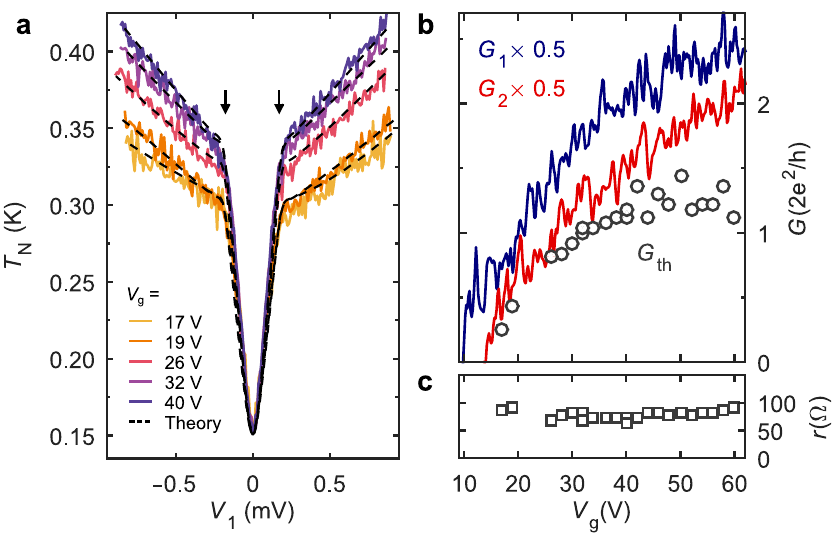}
	\caption{{Thermal conductance in the device NSN-I}. (\textbf{a}) Noise temperature $T_{\mathrm{N}}$ measured in the transmission configuration as {a} function of bias (solid lines, same data as in the lower part of Figure~\ref{fig2}b) along with the model fits (dashed lines). (\textbf{b},\textbf{c}) (symbols)~ Sub-gap thermal conductance $G_{\mathrm{th}}$ and interface resistance parameter $r$ plotted as a function of $V_\mathrm{g}$. (lines) Linear response conductances of the left/right ($G_\mathrm{1/2}$) NS~junctions.}
	\label{fig3}
\end{figure}

We proceed with a quantitative description of the Andreev heat guiding by solving the diffusion equation for the electronic energy distribution (EED), inspired by a quasiclassical approach~\cite{PhysRevB.63.081301, Bubis_2021}. In~the proximitized region, the boundary conditions take into account ARs for the sub-gap transport and residual normal reflections above the gap. Thermal conductance $G_\mathrm{th}$ and interface resistance $r$ are the only two parameters that, together with known $G_1,~G_2$, determine the solution for the EED and the noise temperature $T_\mathrm{N}$ of the floating NS junction~\cite{NAGAEV1992103}.
{For convenience, we choose electrical units for the thermal conductance~\cite{Bubis_2021} $G_\mathrm{th} =  e^2 \nu^*D^*/L_\mathrm{S}$, where $\nu^*$ is the effective one-dimensional density of states, $D^*$ is the diffusion coefficient in the NW region covered by the superconductor and $L_\mathrm{S}$ is the length of the S-terminal. With~this choice, in~case of energy-independent $G_\mathrm{th}$, one can express the heat flux caused by a small thermal bias $\delta T$ applied across the proximity region as $\dot{Q} = G_\mathrm{th}{\cal L}_0 T\delta T $, where ${\cal L}_0 = \pi^2k_\mathrm{B}^2/3e^2$ is the Lorenz number.} The details of theoretical modeling can be found in the Supplemental Materials. In~Figure~\ref{fig3}a, we compare the $T_\mathrm{N}$ measured in the experiment of Figure~\ref{fig2}b (solid lines) with the model fits (dashed lines), {where $T_\mathrm{N}\equiv S_\mathrm{T}/4k_\mathrm{B}G_{2}$}. Plotted as a function of $V_1$ the kinks in $T_\mathrm{N}$ indeed occur at the gap edges for all $V_\mathrm{g}$ values, see the vertical arrows. The~data are perfectly reproduced, ensuring that our model captures correctly the physics of the Andreev heat guiding effect. The~$V_\mathrm{g}$ dependence of the interface parameter $r$ is shown in Figure~\ref{fig3}c. We find $r\sim50~\Omega$, which is consistent with $r_{21}$ in the same device in the normal state, cf. Figure~\ref{fig1}d, and~almost independent of $V_\mathrm{g}$. The~evolution of $G_\mathrm{th}$ at increasing $V_\mathrm{g}$ is shown by symbols in Figure~\ref{fig3}b. The~initial growth is followed by saturation at $G_\mathrm{th}\sim2e^2/h$. This is in contrast with a monotonic increase of the electrical conductances $G_1$,~$G_2$ of NS junctions in the same device, see the lines in Figure~\ref{fig3}b. We attribute this difference to the impact of superconducting proximity effect that diminishes the density of states stronger at higher carrier densities. Note that while the back-gate sensitivity of $G_\mathrm{th}$ is consistent with the behavior of the sub-gap states in the NW region covered {on top} by the superconductor~\cite{Das_2012}, the~microscopic origin of such states and its possible relation, e.g.,~to the spin-orbit coupling in InAs, goes beyond the scope of the present~experiment.

\section{Results: Non-Equilibrium DC~Transport}

\begin{figure}[h]
		\includegraphics[width=13.5 cm]{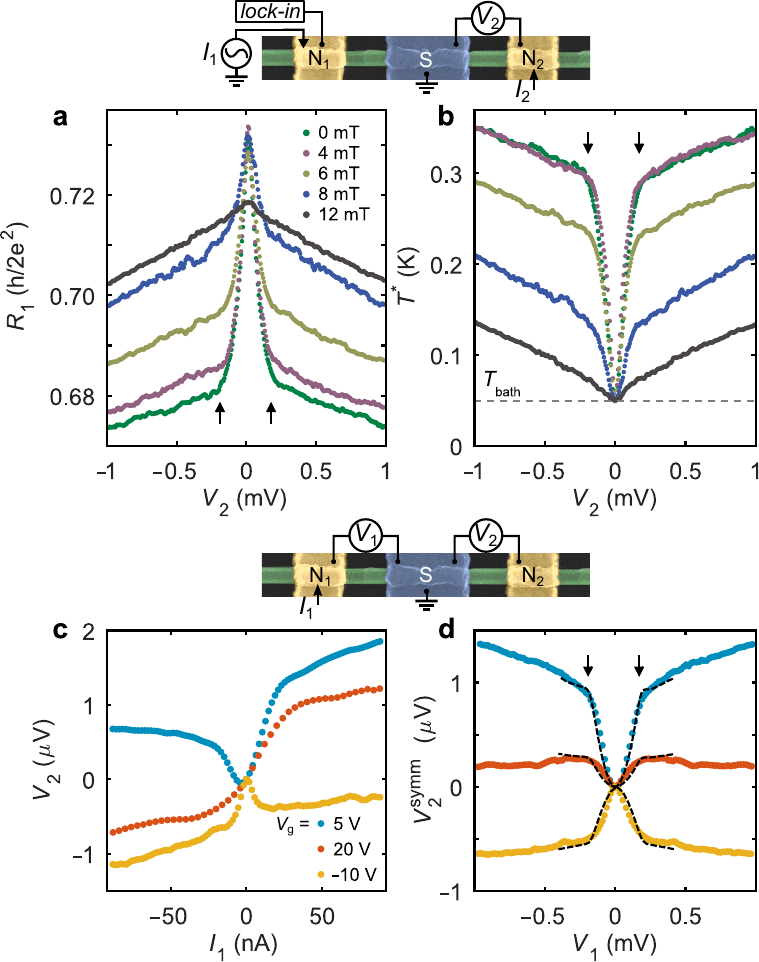}
	\caption{{Resistive thermometry and non-local $I$-$V$s in device NSN-II.} (\textbf{a})~Linear response resistance of the floating NS junction as a function of bias in the neighboring junction. (\textbf{b})~{The s}ame data converted {to} the effective temperature $T^*$. (\textbf{c}) The non-local $I$-$V$ characteristics measured at three representative $V_\mathrm{g}$ values. (\textbf{d}) Symmetric component of the non-local $I$-$V$s. The~dashed lines are the calculated thermoelectric voltage values for different energy-independent Seebeck coefficients of $S/T=3.0~\mu\mathrm{V/K^2},~0.9~\mu\mathrm{V/K^2}$ and $-3.6~\mu\mathrm{V/K^2}$ (from top to bottom). Upper sketch: setup for resistive thermometry. Lower sketch: setup for non-local $I$-$V$s.}
	\label{fig4}
\end{figure}

{So far, we have used shot noise measurements to demonstrate sub-gap Andreev heat guiding. In~the following, we concentrate on the signatures of this effect in charge transport measurements in the device NSN-II. First, we focus on resistive thermometry based on a weak $T$-dependence of the mesoscopic conductance fluctuations. In~Figure~\ref{fig4}a we plot {the} {out-of-equilibrium} linear response resistance $R_1=\partial V_1/\partial I_1|_{I_1=0,I_2\neq0}$ of the floating NS junction as a function of $V_2$ (see the upper sketch in Figure~\ref{fig4} for the measurement configuration). $R_1$ exhibits the same qualitative behavior as the $S_\mathrm{T}$ before, with~much stronger dependence at sub-gap energies, kinks at the gap edges and suppression in $B$-field. Using the equilibrium dependencies $R_1(T)$ for calibration, we converted these data in the effective temperature $T^*$ of the floating NS junction and plotted in Figure~\ref{fig4}b. The~behavior of $T^*$ is similar to that of the $T_\mathrm{N}$ in the device NSN-I, cf. Figure~\ref{fig3}a, potentially making this approach an alternative for the detection of transmitted heat fluxes. Note, however, that resistive thermometry slightly underestimates the effect compared to a simultaneously measured $T_N$, see Supplemental Materials for the details of the analysis. This may be a result of dephasing that causes averaging of the conductance fluctuations and was not taken into~account.}

Finally, we investigate non-local $I$-$V$ characteristics in the configuration shown in the lower sketch of Figure~\ref{fig4}. In~Figure~\ref{fig4}c, the voltage $V_2$ is plotted as a function of $I_1$ for three representative values of $V_\mathrm{g}$. All traces lack full antisymmetry, $V_2(I_1)\neq -V_2(-I_1)$, moreover, the~lower and upper traces exhibit local extrema near the origin, meaning that here the symmetric component dominates the $I$-$V$. This is a signature of the Andreev rectification effect~\cite{PhysRevB.97.045421}, which also caused the asymmetry and sign reversal of $r_{21}$ in Figure~\ref{fig1}d. Figure~\ref{fig4}d shows the symmetric component of the non-local voltage $V_2^\mathrm{symm}\equiv [V_2(I_1)+V_2(-I_1)]/2$ {against} $V_1$. $V_2^\mathrm{symm}$ evolves concurrently to the $T^*$ and $T_N$ with pronounced sub-gap behavior and kinks at $V_1\approx\pm\Delta/e$, see vertical arrows. The~signal is small, in~$1~\mu$V range, with~both the sign and magnitude demonstrating strong $V_\mathrm{g}$-dependent fluctuations, in~contrast with $T^*$ and $T_\mathrm{N}$. We suggest that the finite $V_2^\mathrm{symm}$ has {a} thermoelectric-like origin, analogous to thermopower in Andreev interferometers~\cite{Eom_1998}, and~results from {the} thermal gradient that builds up in response to the transmitted heat flux. More rigorously, in~the absence of inelastic processes in the present experiment, one should think in terms of a spatial gradient of a non-equilibrium EED~\cite{Bubis_2021}. The~data in Figure~\ref{fig4}d are consistent with $V_\mathrm{g}$ fluctuations of the Seebeck coefficient in InAs NWs without superconductors~\cite{Wu_2013,Tikhonov_2016SST} in the range $|S/T|\sim5~\mu\mathrm{V/K^2}$, corresponding fits shown by the dashed lines (see Supplemental Materials for the details). In~the present experiment, thermoelectric-like response also comes from the energy dependence of the mesoscopic fluctuations, but~it can be additionally affected by the Andreev scattering~\cite{PhysRevB.97.045421}. Note that the degree of asymmetry of the non-local conductance $G_{21}\propto-(dV_2/dI_1)$ caused by this effect (see Figure~\ref{fig1}d) is comparable to the data in a Cooper pair splitter~\cite{Hofstetter_2011} and in a tunnel-coupled {Majorana} device~\cite{PhysRevLett.124.036802,puglia2020closing}. Our thermoelectric interpretation may also be useful in explaining these~data.

\section{Discussion} 

{Our experiment reveals the heat-mode excitation in a proximity superconductor via different experimental signatures. On~the one hand, in~DC transport, both in the resistive thermometry (Figure~\ref{fig4}a,b) and in the non-local Andreev rectification (Figure~\ref{fig4}d), the~heat-mode non-equilibrium manifests itself through the energy dependence of sub-gap quasiparticle transmission probabilities. These energy dependencies are encoded in the $T$-dependence of the linear-response diagonal elements of the conductance matrix (see the Supplemental Materials for the details) and in the effective Seebeck coefficient. On~the other hand, in~shot noise, the~energy dependence is irrelevant and the data of Figure~\ref{fig3}a are perfectly fitted with the energy-independent $G_\mathrm{th}$. This difference between the transport and noise approaches is conceptual and lies in the charge-neutral origin of the heat-mode excitation, earlier discussed in Ref.~\cite{Denisov_SST2021}. Below,~we briefly analyze the origin of various non-local responses in the present experiment.}

{Consider for simplicity the case of a single mode NSN device, for~which the non-local electrical and thermal conductances are given by $G_{21} = G_0\mathcal{T}_{21}^-$ and $G_\mathrm{th} = G_0\mathcal{T}_{21}^+$, where $G_0=2e^2/h$ and $\mathcal{T}_{21}^\pm=T_{21}^{he}\pm T_{21}^{ee}$ denote the sum/difference of the non-local Andreev and normal transmission probabilities. The~observation of  $G_\mathrm{th}\gg G_{21}$ implies a predominance of the heat-mode excitation over the charge-mode, that is $T_{21}^{he}\approx T_{21}^{ee}\gg |\mathcal{T}_{21}^-|$. In~this situation, a~weak energy dependence of the transmission probabilities  primarily affects the $G_{21}$. Within~the first-order expansion $\mathcal{T}_{21}^-=\mathcal{T}_{21}^-(0)+\varepsilon \left(d\mathcal{T}_{21}^-/d\varepsilon\right)$, therefore, the non-local $I-V$ characteristics acquire symmetric component. 
Using the formalism of Ref.~\cite{Anantram_1996}, we obtain for the configuration of the bottom sketch in Figure~\ref{fig4}: $V_2^\mathrm{symm} = -|e|(G_0/G_{22})\left(d\mathcal{T}_{21}^-/d\varepsilon\right)(V_1)^2/2$, or, equivalently, $V_2^\mathrm{symm} = -|e|(dG_{21}/d\varepsilon)(V_1)^2/2G_{22}$.  The latter relation is also valid in the multimode case, bridging the effective Seebeck coefficient with the energy dependence of the spectral conductance. Similarly, the~energy dependence of the diagonal conductance $G_{22}(\varepsilon)$ is responsible for the resistive thermometry signal in the configuration of the top sketch in Figure~\ref{fig4}. Here, the~non-zero term comes from the second derivative $d^2G_{22}/d\varepsilon^2$, as~follows from the derivation given in the Supplemental Materials. Such effects are completely irrelevant for the non-local shot noise measurement in the transmission configuration. Estimated from Figure~\ref{fig4}d, the~energy dependence of the transmission probabilities can result in $\sim1\%$ variation of the $G_\mathrm{th}(\varepsilon)$ within the sub-gap window $|\varepsilon|<\Delta$ in the device NSN-II. Hence, $G_\mathrm{th}(\varepsilon)\approx const$ and the shot noise in the transmission configuration reads $S_\mathrm{T} = 2|eV_1|G_\mathrm{th}$ (at $T=0$). Note, however, that the energy-independent $G_\mathrm{th}$ is puzzling itself and, obviously, contradicts the expected presence of the induced superconducting gap in the proximitized NW region. A~microscopic resolution of this puzzle is a difficult theoretical task and goes beyond the scope of the present work.} 

In summary, we investigated the heat-mode excitation manifesting itself in various non-local responses in NSN proximity devices based on InAs NWs.  In~DC transport, the~non-local signals couple to the heat-mode only indirectly, via a weak and non-universal energy dependence of the spectral conductance. This is in stark contrast with our shot noise approach, which senses the randomness caused by the non-equilibrium EED itself, without~the need for any type of spectral resolution~\cite{tikhonov2020energy}. In~the same way, the~shot noise can also probe excitations of different origin, e.g.,~spin currents in superconducting spintronics~\cite{Linder_2015}, or~even valley currents~\cite{Schaibley_2016}, by~virtue of spontaneous fluctuations that arise when such currents are fed into the adjacent normal lead~\cite{PhysRevB.84.073302,Arakawa_2015,Khrapai_2017,PhysRevResearch.2.023221}. {Possible applications are not at all limited to the NW-based material platforms}. From~this perspective, our experiment establishes a natural background to probe charge-neutral excitations, both above-gap in bulk superconductors and sub-gap in proximity superconductors, including the proposed detection of Majorana zero modes in heat transport~\cite{PhysRevB.61.10267,Zhang_thermal_2011,PhysRevLett.106.057001,PhysRevLett.117.196801,Kasahara2018} and, possibly, in measurements of the entanglement entropy~\cite{Muralidharan2022}.

\vspace{6pt} 


{\textbf{Supplemetary.} The Supplemental Materials for this article contain Figure S1: Sketch of the experimental setup; Figure S2: Calibration via equilibrium noise; Figure S3: Shot-noise analysis; Figure S4: Additional data in device NSN-I: local conductance; Figure S5: Additional data in device NSN-II: local conductance; Figure S6: Additional data in device NSN-I: non-local conductance; Figure S7: Additional data in device NSN-II: non-local conductance; Figure S8:. Effective resistance model for NW/S interface; Figure S9: T-dependence in the linear response regime and calibration of the resistive thermometry; Figure S10: T-dependence beyond the linear response regime; Figure S11: Analytical model: layout and EED; Figure S12: Analytical model: results; Figure S13: Comparison of the non-local noise thermometry and resistive thermometry; Figure S14: Superconducting critical temperature of the Al-film. Supplemental Materials cite References~\cite{PhysRevB.63.081301,Tikhonov2016,NAGAEV1992103,BLANTER20001,Bubis_2017,Hertenberger_2010,PhysRevB.97.115306}.} 

{\textbf{Author Contributions.} Shot noise experiments: A.D. and S.P.; 
transport experiments: A.D., A.B. and S.P.; fabrication: A.B. and N.T.; nanowire growth: J.B., J.T., D.R. and G.K.; 
modeling: A.D. and S.P.; supervision, writing and project administration: E.T., A.N. and V.K. All authors have read and agreed to the published version of the~manuscript.}

{\textbf{Funding.} Implementation of resistive thermometry and its comparison to noise thermometry was supported by the Russian Science Foundation Grant No. 18-72-10135. Theoretical modeling was performed under the state task of the ISSP RAS.}

{\textbf{Data Availability.} The full data for this study can be obtained from the corresponding author upon reasonable request.} 

{\textbf{Acknowledgements.} We are grateful to S.M. Frolov, {A.P. Higginbotham,} T.M. Klapwijk, S. Ludwig,  A.S. Mel'nikov  and K.E. Nagaev for helpful~discussions.}

\end{document}


\title{Heat-mode excitation in a proximity superconductor. Supplemental Materials}
\author{A.O.~Denisov}
\affiliation{Osipyan Institute of Solid State Physics, Russian Academy of Sciences, 142432 Chernogolovka, Russia}
\affiliation{Department of Physics, Princeton University, Princeton, New Jersey 08544, USA}
\author{A.V.~Bubis} 
\affiliation{Skolkovo Institute of Science and Technology, Nobel street 3, 121205 Moscow, Russian Federation}
\affiliation{Osipyan Institute of Solid State Physics, Russian Academy of Sciences, 142432 Chernogolovka, Russia}
\author{S.U.~Piatrusha}
\affiliation{Osipyan Institute of Solid State Physics, Russian Academy of Sciences, 142432 Chernogolovka, Russia}
\author{N.A.~Titova}
\affiliation{Moscow Pedagogical State University, 29 Malaya Pirogovskaya St, 119435 Moscow, Russia}
\author{A.G.~Nasibulin}
\affiliation{Skolkovo Institute of Science and Technology, Nobel street 3, 121205 Moscow, Russian Federation}
\author{J.~Becker}
\affiliation{Walter Schottky Institut, Physik Department, and Center for Nanotechnology and Nanomaterials, Technische Universit\"{a}t M\"{u}nchen, Am Coulombwall 4, Garching 85748, Germany}
\author{J.~Treu}
\affiliation{Walter Schottky Institut, Physik Department, and Center for Nanotechnology and Nanomaterials, Technische Universit\"{a}t M\"{u}nchen, Am Coulombwall 4, Garching 85748, Germany}
\author{D.~Ruhstorfer}
\affiliation{Walter Schottky Institut, Physik Department, and Center for Nanotechnology and Nanomaterials, Technische Universit\"{a}t M\"{u}nchen, Am Coulombwall 4, Garching 85748, Germany}
\author{G.~Koblm\"{u}ller}
\affiliation{Walter Schottky Institut, Physik Department, and Center for Nanotechnology and Nanomaterials, Technische Universit\"{a}t M\"{u}nchen, Am Coulombwall 4, Garching 85748, Germany}
\author{E.S.~Tikhonov}
\affiliation{Osipyan Institute of Solid State Physics, Russian Academy of Sciences, 142432 Chernogolovka, Russia}
\author{V.S.~Khrapai}
\affiliation{Osipyan Institute of Solid State Physics, Russian Academy of Sciences, 142432 Chernogolovka, Russia}
\affiliation{National Research University Higher School of Economics, 20 Myasnitskaya Street, 101000 Moscow, Russia}

\begin{abstract}
	This file contains supplementary information for the main text, including the following figures: \\
	Figure~\ref{sup_fig5}: Sketch of the experimental setup \\
	Figure~\ref{sup_fig11}: Calibration via equilibrium noise  \\
	Figure~\ref{sup_fig12}: Shot-noise analysis  \\
	Figure~\ref{sup_fig6}: Additional data in device NSN-I: local conductance  \\
	Figure~\ref{sup_fig8}: Additional data in device NSN-II: local conductance  \\
	Figure~\ref{sup_fig7}: Additional data in device NSN-I: non-local conductance  \\
	Figure~\ref{sup_fig9}: Additional data in device NSN-II: non-local conductance   \\
	Figure~\ref{sup_fig1}:. Effective resistance model for NW/S interface  \\
	Figure~\ref{sup_fig2_new}: T-dependence in the linear response regime and calibration of the resistive thermometry  \\
	Figure~\ref{sup_fig10}: T-dependence beyond the linear response regime  \\
	Figure~\ref{sup_fig4}: Analytical model: layout and EED  \\
	Figure~\ref{fig:T_vs_R}: Analytical model: results  \\
	Figure~\ref{fig:t_n_t_star}: Comparison of the non-local noise thermometry and resistive thermometry  \\
	Figure~\ref{sup_fig_al}: Superconducting critical temperature of the Al-film
\end{abstract}

\maketitle

	
	\section*{Noise and charge transport measurements}
	
	\begin{figure*}[h]
		\begin{center}
			\includegraphics[scale=0.7]{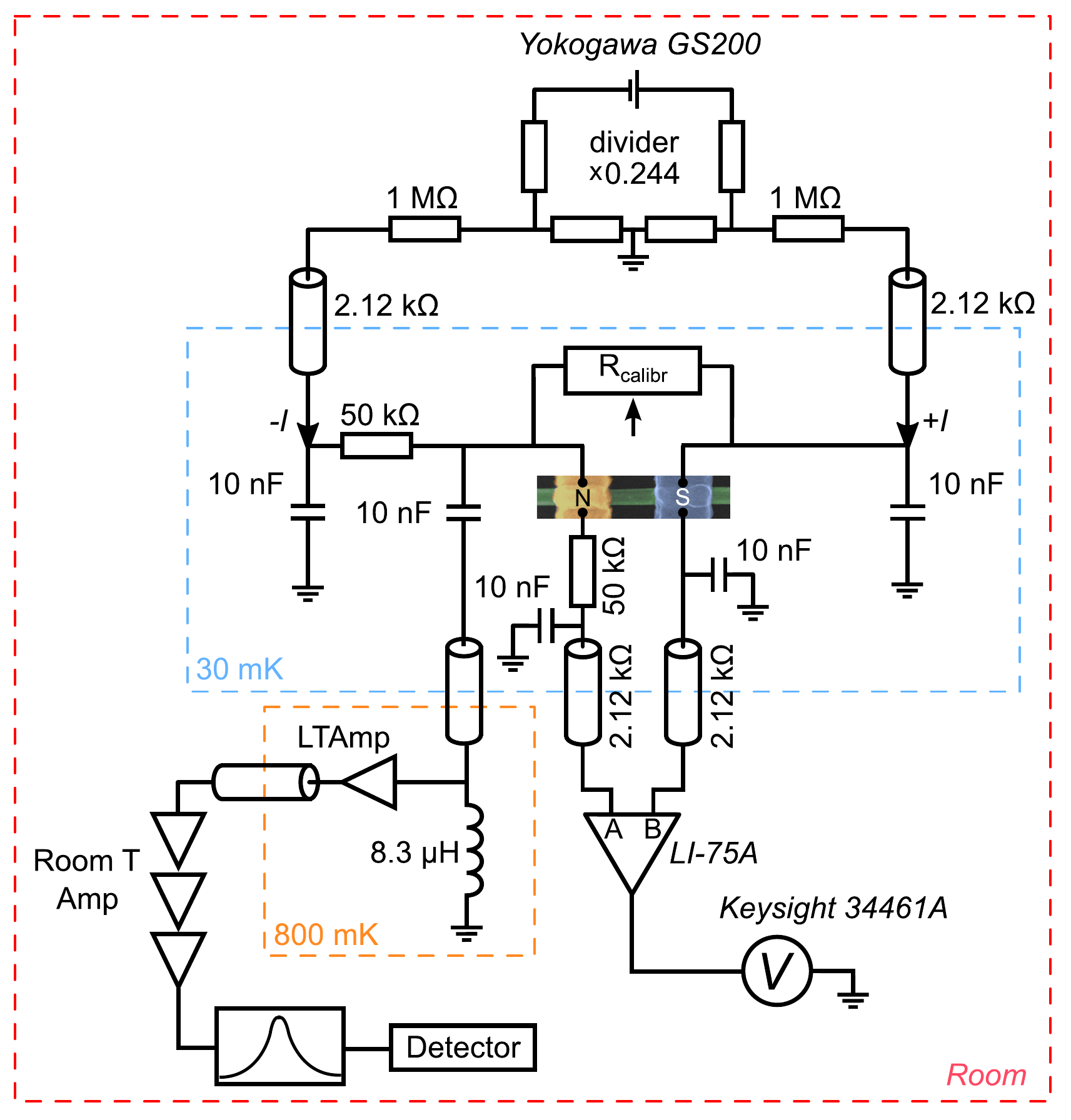}
		\end{center}
		\caption{\textbf{Sketch of the experimental setup.} }
		\label{sup_fig5}
	\end{figure*}
DC and low frequency AC transport measurements are carried out using symmetric current bias scheme with divider shown in Fig.~\ref{sup_fig5}. We use quasi-4-terminal setup thus excluding wiring and filtering contribution into measured voltage signal. We use SR-$7265$ lock-in for resistive thermometry with typical modulation current of $2~\mathrm{nA}$, $f=19.3~\mathrm{Hz}$, time constant = $2\mathrm{s}$, AC gain $30~\mathrm{dB}$ and filter slope of 24~$\mathrm{dB/oct}$.   
	
	The noise spectral density was measured using the home-made low-temperature amplifier (LTamp) with a voltage gain of about \myq{10}{dB} and the input current noise of $\sim 2$--\myq{6\times10^{-27}}{A^2/Hz}. The voltage fluctuations on a \myq{25}{k\Omega} load resistance were measured near the central frequency \myq{14.2}{MHz} (\myq{\pm 0.6}{MHz} for $-3~\mathrm{dB}$ point) of a resonant circuit at the input of the LTAmp. The output of the LTamp was fed into the low noise \myq{75}{dB} gain room temperature amplification stage followed by a hand-made analogue band-pass filter and a power detector. The setup was calibrated using HEMT ATF-35143 as adjustable load $R_{\mathrm{calib}}= 50~\mathrm{\Omega}~\to~>100~\mathrm{M\Omega}$ for the equilibrium Johnson-Nyquist noise thermometry. Except for the periods of calibration, the transistor was always kept pinched off. Unless otherwise stated, the measurements were performed in a cryogenic free Bluefors dilution refrigerator BF-LD250 at a bath temperature of \myq{30}{mK}.
	\newpage
	\clearpage
	\section*{Johnson-Nyquist noise thermometry}
	\begin{figure*}[h]
		\begin{center}
			\includegraphics[scale=1]{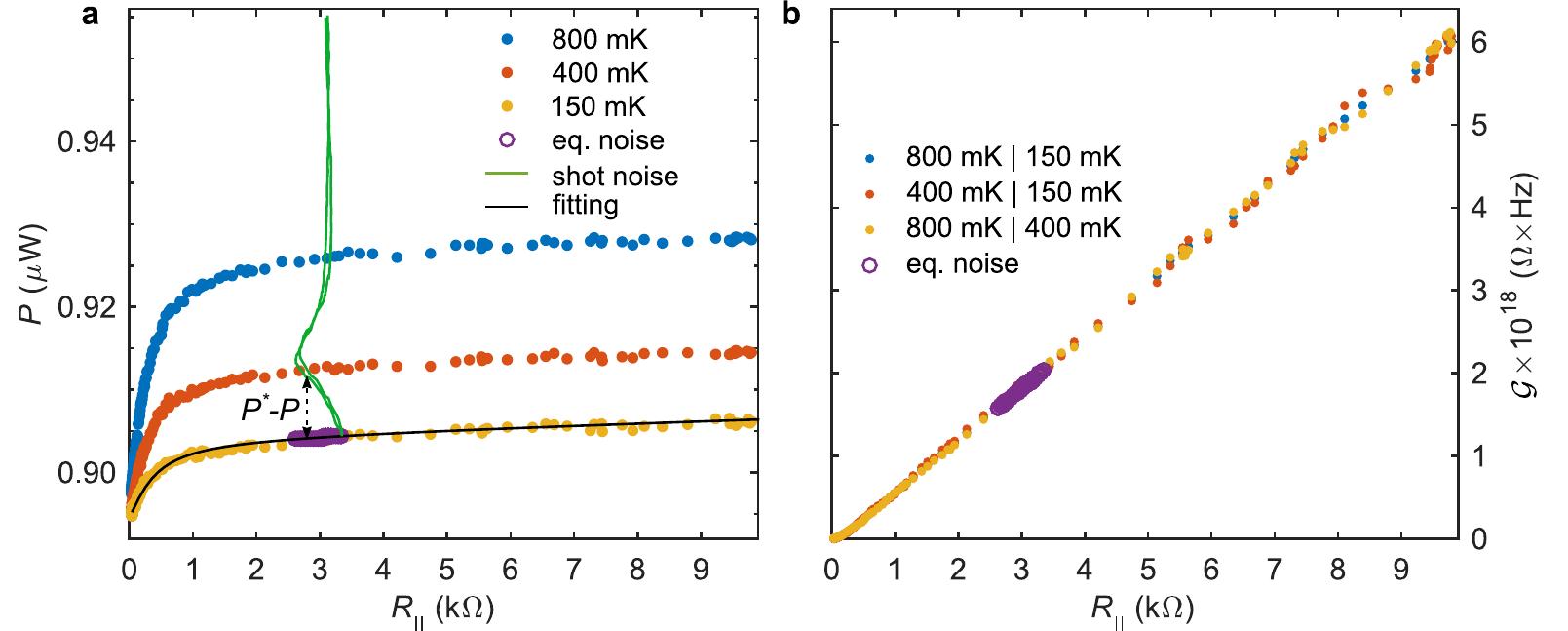}
		\end{center}
	\caption{\textbf{Calibration via equilibrium noise.} (a) Power released on the detector as a function of total load resistance $R_{||}$ measured at different bath temperatures (see electron temperature in the legend). Green solid line shows typical reflected shot-noise signal at applying bias current through the device (NSN~-~I, $V_{\mathrm{g}}=40~\mathrm{V}$). Black solid line is fit using Nyquist relation. Purple circles shows bias-dependent gain and equilibrium noise. (b) Total gain of the setup. Three sets of points correspond to three different combinations of $T_{1}|T_{2}$ (see text below).}
		\label{sup_fig11}
	\end{figure*}
\begin{figure*}[h]
	\begin{center}
		\includegraphics[scale=1]{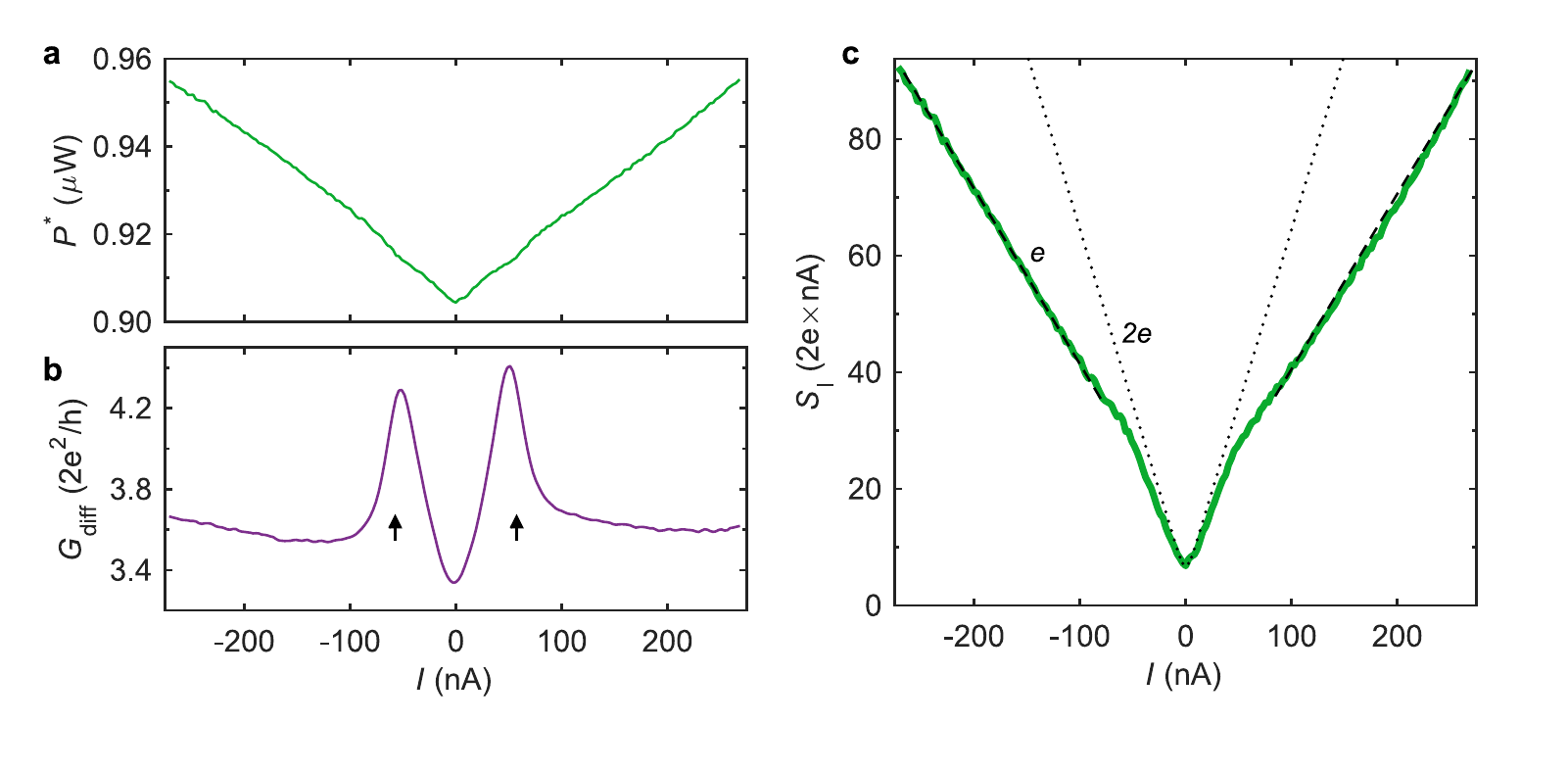}
	\end{center}
	\caption{\textbf{Shot-noise analysis.} (a) Power dissipated in the detector is plotted as a function of current through the sample. (b) Differential conductance of the NS junction. Arrows mark superconducting gap $\pm\Delta$. (c) Extracted current noise spectral density. Dashed line corresponds to Fano-factor $F~=~S_{\mathrm{I}}/2eI~=~0.30$ and $e^{*}=e$, dotted line is for doubled charge - $F~=~0.30$ and $e^{*}=2e$.  }
	\label{sup_fig12}
\end{figure*}
\newpage
We calibrate our setup using equilibrium noise thermometry. At zero current through the device, we are able to change a value of $R_{||}=(G_{\mathrm{diff}}+R_{\mathrm{25k\Omega}}^{-1}+R_{\mathrm{calibr}}^{-1})^{-1}$ drastically, where $G_{\mathrm{diff}}$ is a differential resistance (zero bias) of the sample shown in Fig.~\ref{sup_fig12}b. Power released on the detector after all amplification stages :
	\begin{equation}
	P(R_{||})=\Big(\frac{4k_{\mathrm{B}}T}{R_{||}}+S_{\mathrm{I}}^{\mathrm{Amp}}\Big)\int\frac{G\times Tr^{\mathrm{filter}}(f)}{R_{||}^{-2}+|Z_{\mathrm{LC}}|^{-2}}df+P_{0}=\mathcal{G}(R_{||})\Big(\frac{4k_{\mathrm{B}}T}{R_{||}}+S_{\mathrm{I}}^{\mathrm{Amp}} \Big)+P_{0}
	\end{equation}
	where $G$ is an unknown total gain, $Z_{\mathrm{LC}}$ - complex impedance of the $\mathrm{LC}$ contour, $Tr^{\mathrm{filter}}(f)$ - transmission characteristic of the band-pass filter, $S_{\mathrm{I}}^{\mathrm{Amp}}$ and $P_{0}$ - parasitic current noise and background of the low-temperature amplifier. After an integration over frequency, we can use generalized value for gain $\mathcal{G}(R_{||})$ which can be extracted by measuring $P(R_{||})$ at different bath temperatures:
	\begin{equation}
	\mathcal{G}(R_{||})=\frac{P(R_{||},T_{1})-P(R_{||},T_{2})}{T_{1}-T_{2}}\frac{R_{||}}{4k_{\mathrm{B}}}
	\end{equation}

When we apply current through the sample, the crossover from thermal to non-equilibrium shot noise (see solid green curve in Fig.~\ref{sup_fig11}a and Fig.~\ref{sup_fig12}a) appears. Depending on the bias current, $R_{||}(I)$ is changing thus making $\mathcal{G}(R_{||})$ bias-dependent (see purple symbols in Fig.~\ref{sup_fig11}a,~b). Desired current noise of the sample $S_{\mathrm{I}}$ contributes to the total power as follows (transistor is pinched off $R_{\mathrm{calib}}>100~\mathrm{M\Omega}$) :
	\begin{equation}
P^{*}(R_{||})=\mathcal{G}(R_{||})\Big(\frac{4k_{\mathrm{B}}T}{R_{25\mathrm{k\Omega}}}+S_{\mathrm{I}}+S_{\mathrm{I}}^{\mathrm{Amp}} \Big)+P_{0},~~S_{\mathrm{I}}=\frac{P^{*}(R_{||})-P(R_{||})}{\mathcal{G}(R_{||})}+{4k_{\mathrm{B}}T}{G_{\mathrm{diff}}}
	\end{equation}
Finalized current spectral density curve is shown in Fig.~\ref{sup_fig12}c.

	\newpage
\clearpage
	\section*{Local charge transport}
	\begin{figure*}[h]
		\begin{center}
			\includegraphics[scale=1]{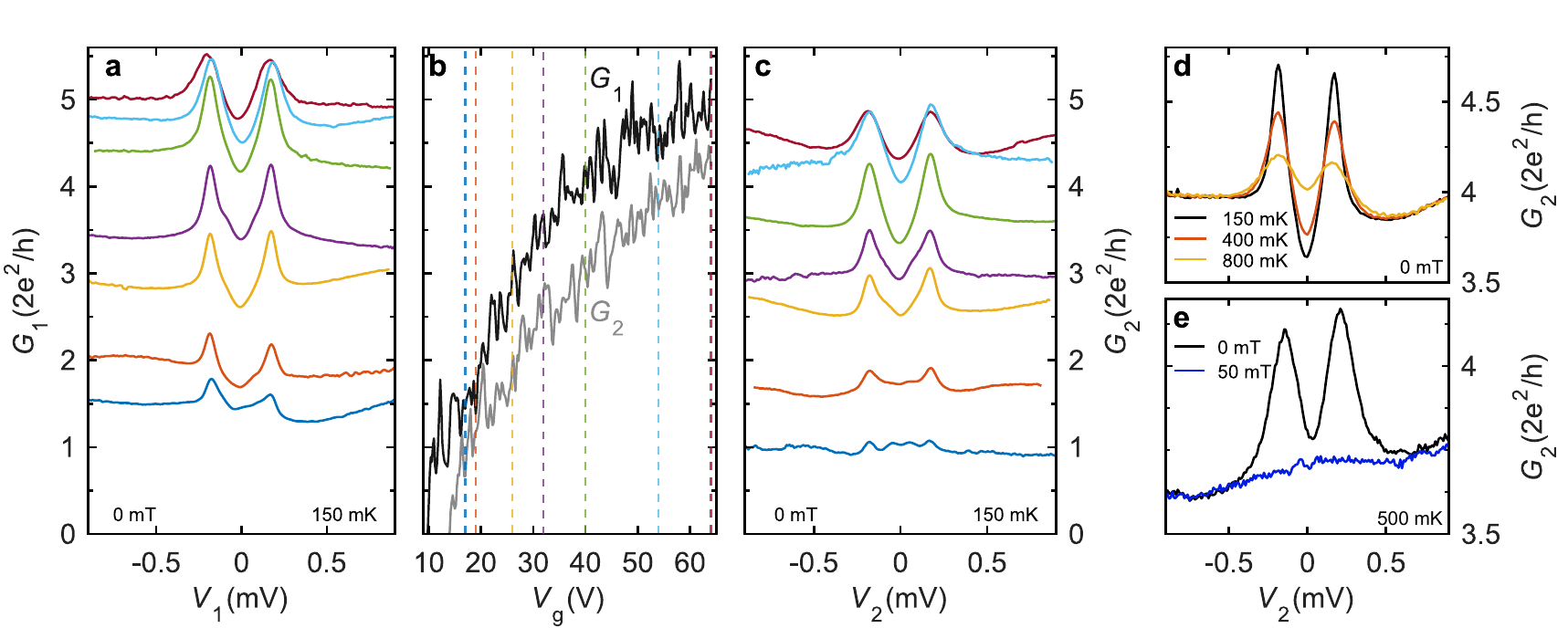}
		\end{center}
		\caption{\textbf{Additional data in device NSN-I: local conductance.} (a) and (c) Local spectral conductance of the left and right NS junctions correspondingly, measured at different back-gate voltages $V_{\mathrm{g}}=~17,~19~,26,~32,~40,~54,~64~\mathrm{V}$ from bottom to top. (b) Linear-response conductance is plotted as function of $V_{\mathrm{g}}$. Dashed lines of corresponding colors point certain values of back-gate voltages from (a) and (c). (d) and (e) Temperature and magnetic field dependence of the spectral conductance measured at constant $V_{\mathrm{g}}=41$ and $50~\mathrm{V}$ correspondingly.}
		\label{sup_fig6}
	\end{figure*}
	\begin{figure*}[h]
		\begin{center}
			\includegraphics[scale=1]{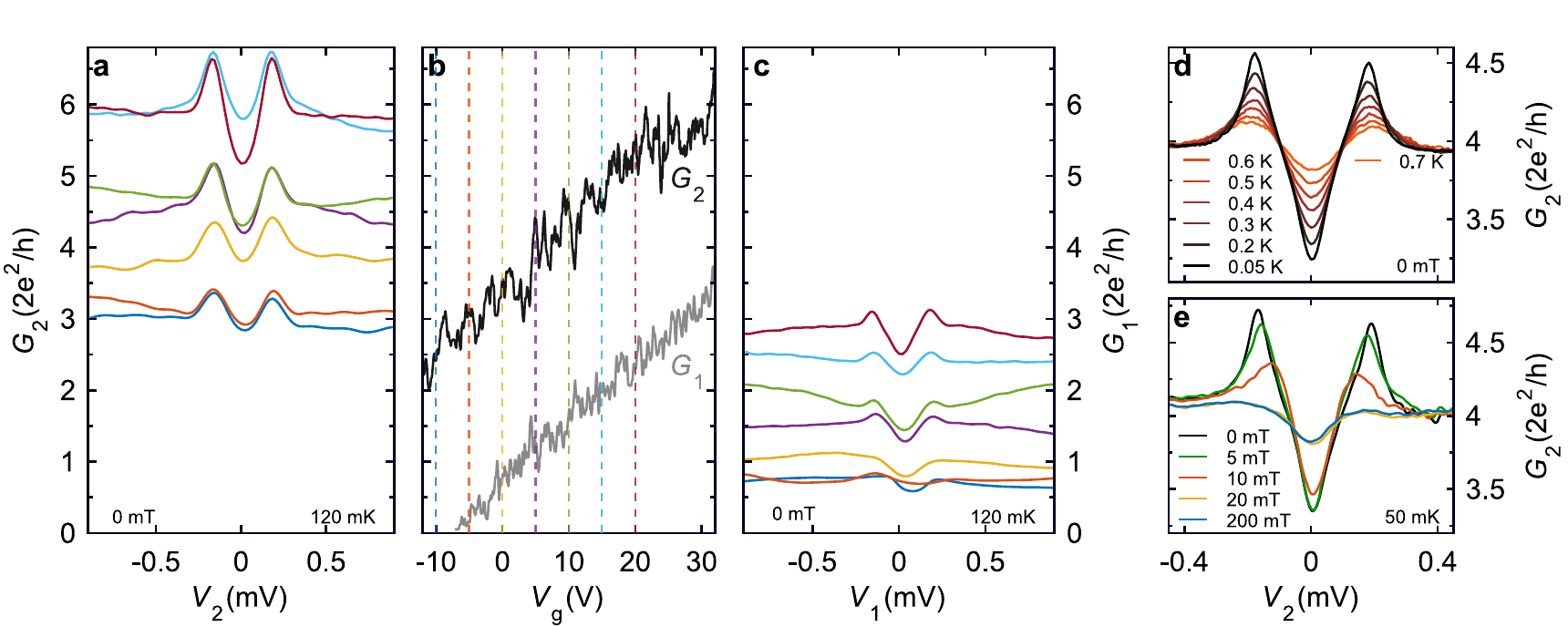}
		\end{center}
		\caption{\textbf{Additional data in device NSN-II: local conductance.} (a) and (c) Local spectral conductance of the left and right NS junctions correspondingly, measured at different back-gate voltages $V_{\mathrm{g}}=~-10,~-5~,0,~5,~10,~15,~20~\mathrm{V}$ from bottom to top. (b) Linear-response conductance is plotted as function of $V_{\mathrm{g}}$. Dashed lines of corresponding colors point certain values of back-gate voltages from (a) and (c). (d) and (e) Temperature and magnetic field dependence of the spectral conductance measured at constant $V_{\mathrm{g}}=0~\mathrm{V}$.}
		\label{sup_fig8}
	\end{figure*}
	
	\newpage
	\section*{Non-local charge transport}
	\begin{figure*}[h]
		\begin{center}
			\includegraphics[scale=0.78]{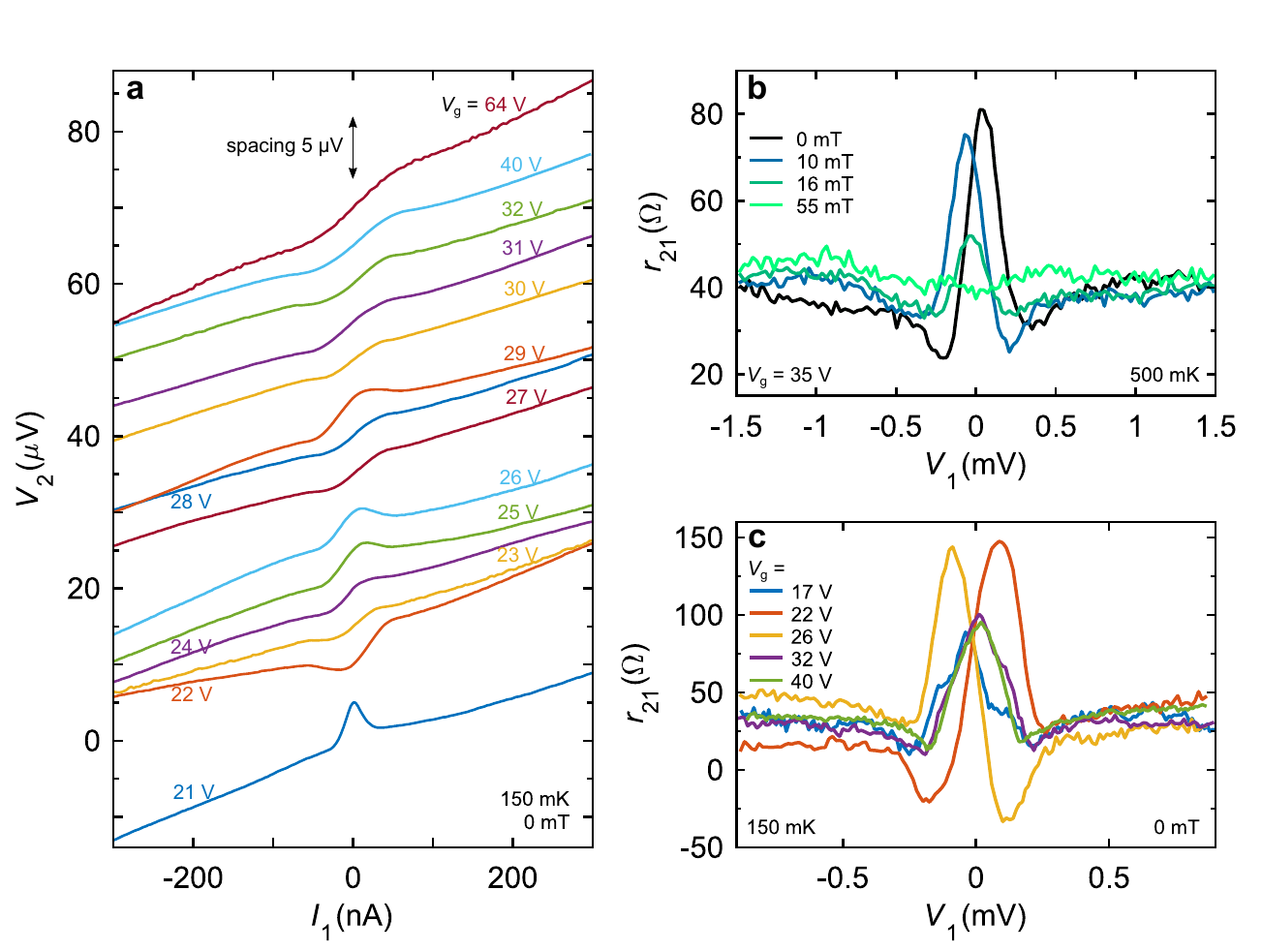}
		\end{center}
		\caption{\textbf{Additional data in device NSN-I: non-local conductance.} (a) Non-local I/V characteristics measured at different back-gate voltages. For convenience, curves are shifted vertically with spacing of $5~\mathrm{\mu V}$. (b) and (c) Differential non-local resistance $r_{21}=dV_{2}/dI_{1},~I_{2}=0$ at different magnetic fields and back-gate voltages correspondingly.}
		\label{sup_fig7}
	\end{figure*}
	\begin{figure*}[h]
		\begin{center}
			\includegraphics[scale=0.78]{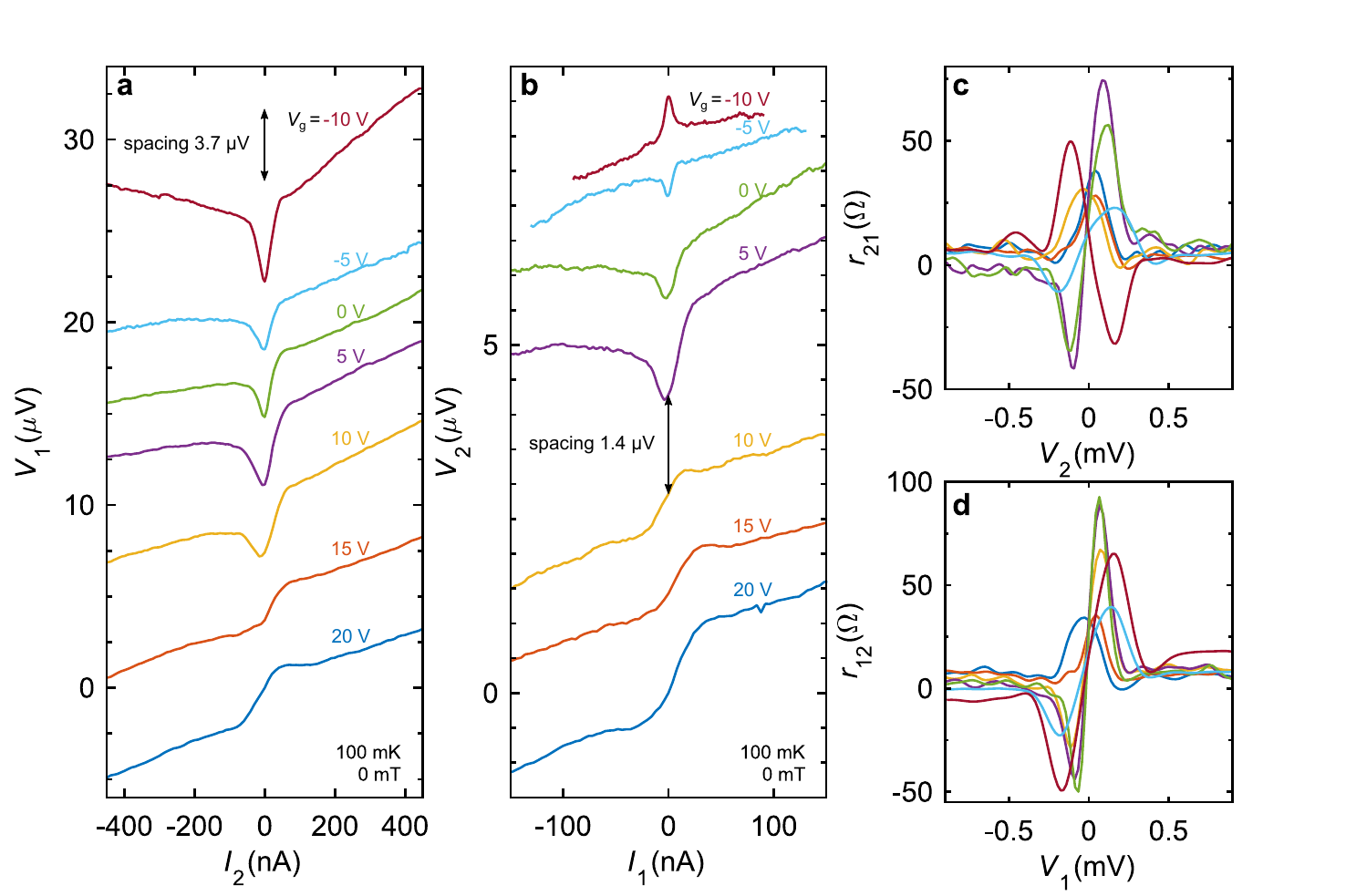}
		\end{center}
		\caption{\textbf{Additional data in device NSN-II: non-local conductance.} (a) and (b) Non-local I/V characteristics of both NS junctions measured at different back-gate voltages. For convenience, curves are shifted vertically with spacing of $3.7~\mathrm{\mu V}$ and $1.4~\mathrm{\mu V}$ correspondingly. (c) and (d) Differential non-local resistance $r_{21}=dV_{2}/dI_{1},~I_{2}=0$ and $r_{12}=dV_{1}/dI_{2},~I_{1}=0$ at different back-gate voltages. Colors match those from (a) and (b).}
		\label{sup_fig9}
	\end{figure*}
	
	\newpage
	\clearpage
	\section*{Current transfer length estimation}
	\begin{figure*}[h]
		\begin{center}
			\includegraphics[scale=1]{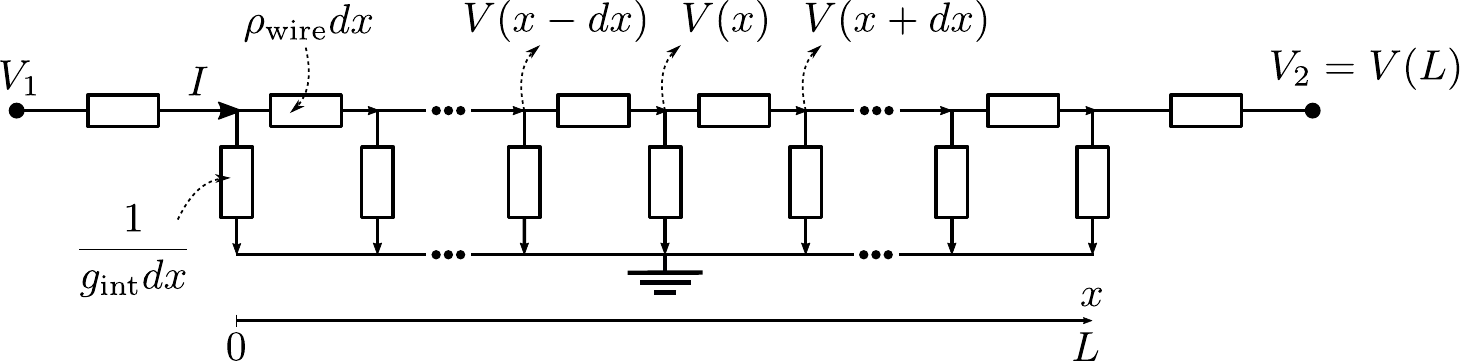}
		\end{center}
		\caption{\textbf{Effective resistance model for nanowire/superconductor interface.} }
		\label{sup_fig1}
	\end{figure*}
	
	To estimate the characteristic length of charge overflow within grounded S terminal $l_{\mathrm{T}}$ we use circuit shown in fig.\,\ref{sup_fig1}. Here $\rho_{\mathrm{wire}}$ and $g_{\mathrm{int}}$ are resistance of the nanowire (NW) and conductivity of interface per unit length respectively. In the continuous limit we can write current conservation for each point along NW/S interface:
	\begin{equation*}
		\begin{gathered}
			\frac{V(x+dx)-V(x)}{\rho_{\mathrm{wire}} dx}+\frac{V(x-dx)-V(x)}{\rho_{\mathrm{wire}} dx}=\frac{V(x)dx}{1/g_{\mathrm{int}}} \\ 
			\frac{d^2V(x)}{dx^2}=\frac{V(x)}{l_{\mathrm{T}}^2},~~l_{\mathrm{T}}=\frac{1}{\sqrt{\rho_{\mathrm{wire}}g_{\mathrm{int}}}}
		\end{gathered}
	\end{equation*}
	Boundary conditions including one that normal terminal N2 is floating and no current flow into it. \begin{equation*}
		\frac{dV(x)}{dx}\bigg{|}_{x=0}=-\rho_{\mathrm{wire}} I,\;\; 
		\frac{dV(x)}{dx}\bigg{|}_{x=L}=0 
	\end{equation*}
	Solving elementary Neumann problem we can find non-local rsistance $r_{21}$:
	\begin{equation*}
		\begin{gathered}
			r_{21}=\frac{V(L)}{I}=\frac{l_{\mathrm{T}}\rho_{\mathrm{wire}}}{\sinh(\frac{L}{l_{\mathrm{T}}})}
		\end{gathered}
	\end{equation*}
	For two measured devices (NSN-I, NSN-II) we have $r_{21}\approx40,~10\,\Omega$ and $L\approx200,~300\,\mathrm{nm}$ respectively, thus $l_{\mathrm{T}}\approx75\,\mathrm{nm}$.
	\newpage
	\section*{Temperature dependence of differential conductance.}
	\begin{figure*}[h]
		\begin{center}
			\includegraphics[scale=1]{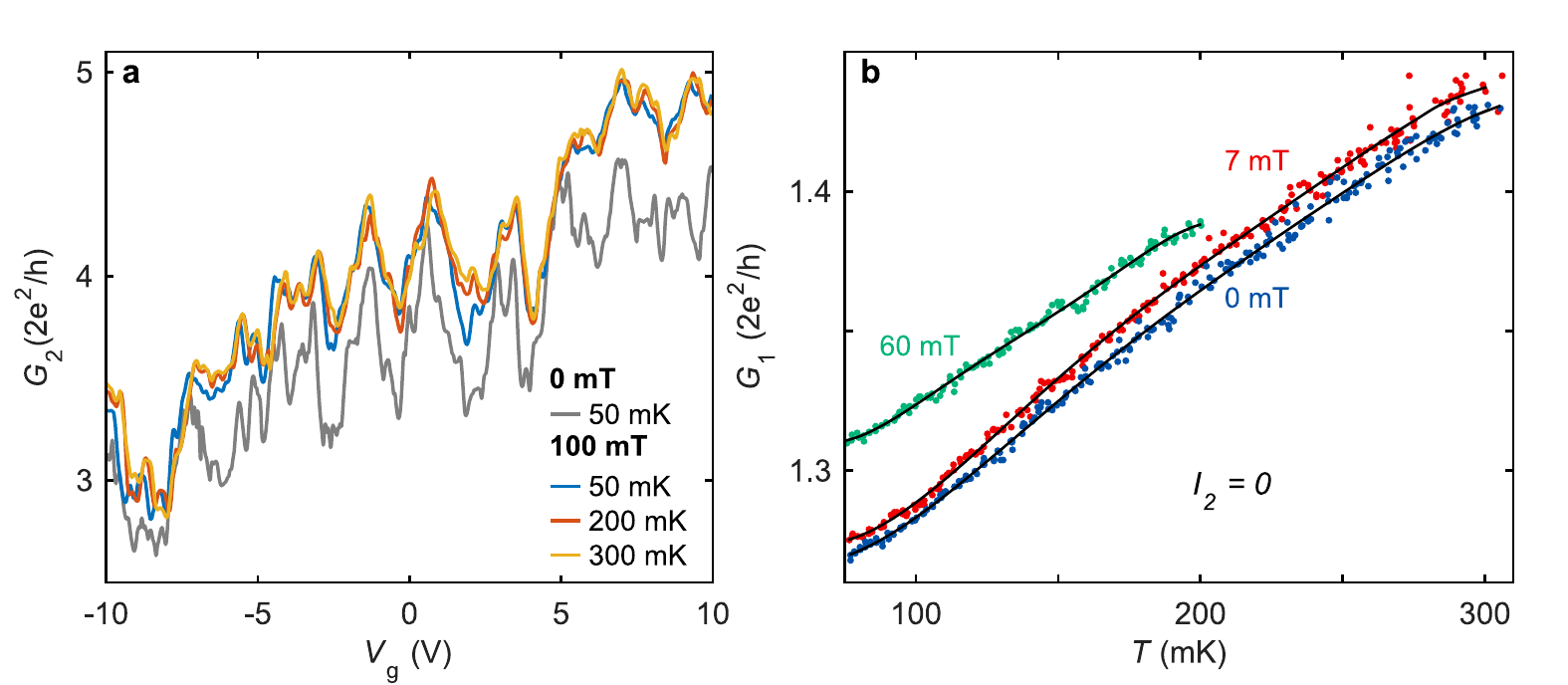}
		\end{center}
		\caption{\textbf{$T$-dependence in the linear response regime and calibration of the resistive thermometry.} (a)~Linear-response conductance of a single NS junction~(NSN~-~II device) measured at different bath temperatures. Overall increasing dependence with reproducible UCF persists up to the shift~$\sim10\%$ when large magnetic field ($100\,\mathrm{mT}>B_{\mathrm{c}}$) is applied. Such zero-bias deep is more evident from differential conductance data measured at $V_{\mathrm{g}}=0\,\mathrm{V}$ and $I_{2}=0$ shown in (b). Solid lines are smooth polynomial fits we use for the resistive thermometry.}
		\label{sup_fig2_new}
	\end{figure*}
	
	\begin{figure*}[h]
		\begin{center}
			\includegraphics[scale=1]{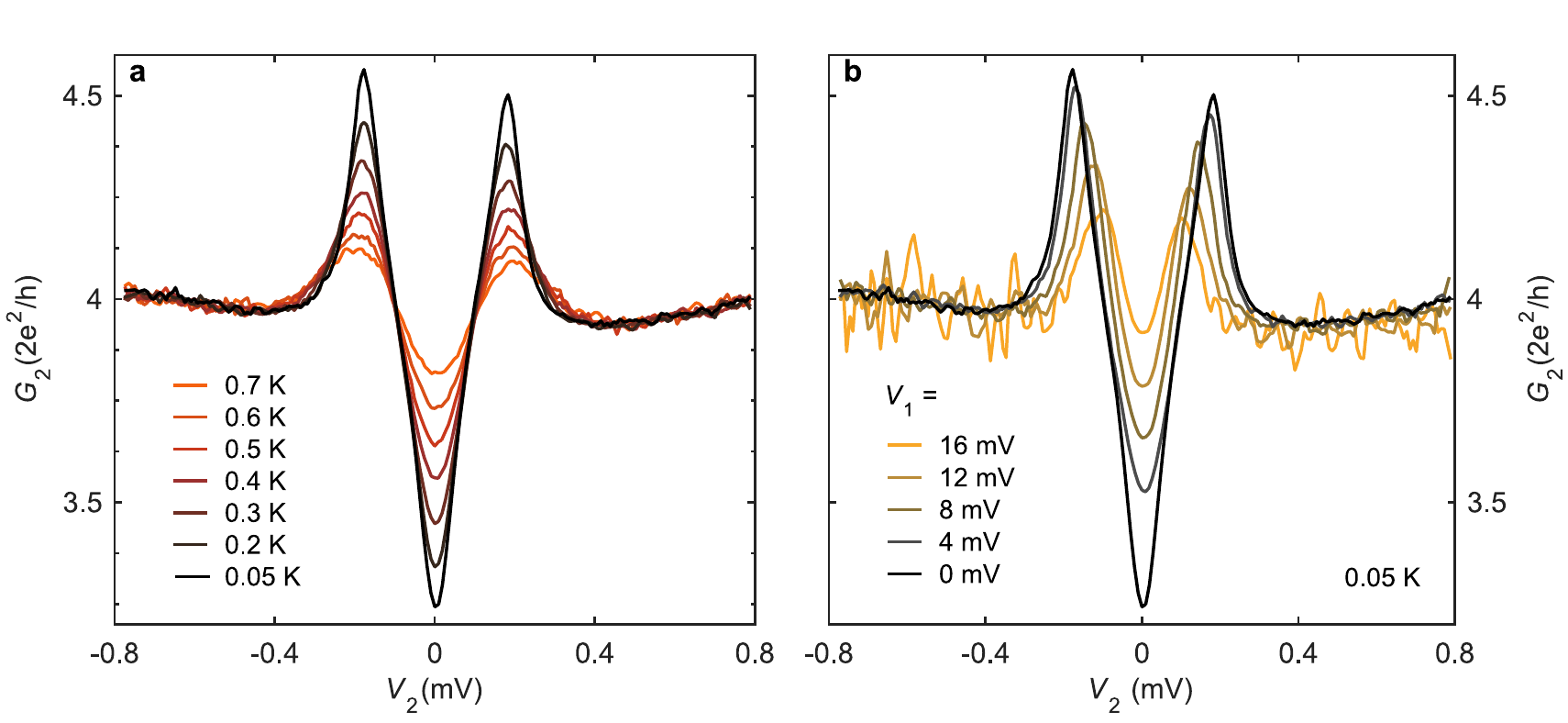}
		\end{center}
		\caption{\textbf{$T$-dependence beyond the linear response regime.} (a) Spectral conductance measured at constant $V_{\mathrm{g}}=~0~\mathrm{V}$ (NSN~-~II device) and different bath temperatures. (b) Spectral conductance measured at constant $V_{\mathrm{g}}=~0~\mathrm{V}$, bath temperature but different "heating" voltages across the adjacent NS junction.}
		\label{sup_fig10}
	\end{figure*}
	
	\newpage

	\section{\textcolor{black}{Analytical model}}
	\begin{figure*}[h]
		\begin{center}
			\vspace{-0.5cm}
			\includegraphics[scale=1]{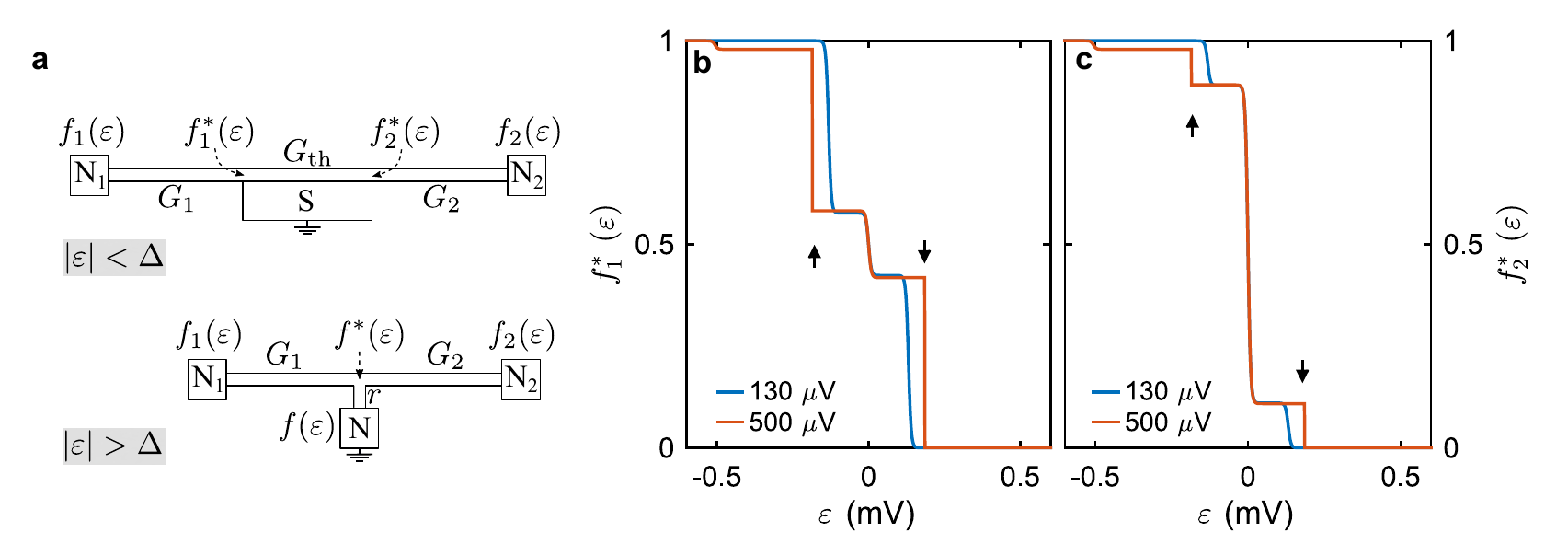}
			\vspace{-0.5cm}
			\caption{\textbf{Analytical model: layout and EED.} (a) Schematics of the analytical model, separately for sub-gap and above-gap quasiparticle energies. (b)~-~(c) Calculated EEDs near two ends of the proximitized sections for $V_{1}=130~\mathrm{\mu V}$ and $500~\mathrm{\mu V}$ ($T=50~\mathrm{mK},~G_{\mathrm{th}}=~1.17\times2e^2/h,~r=64~\Omega$). Arrows mark the superconducting gap of Al, $\Delta=180~\mu V$. }
			\label{sup_fig4}
		\end{center}
	\end{figure*}
	\textcolor{black}{The sketch of analytical model we use to fit experimental data is shown in fig.3a. Since we operate in the non-linear $\varepsilon_{T}\sim k_{\mathrm{B}}T\ll eV$ regime, where $\xi_{T}=\hbar D/L^2\approx 15\,\mathrm{\mu V}$ is the Thouless energy, we can neglect the penetration of the condensate into NW and consider it just as a normal metal with a non-equilibrium electronic energy distribution (EED). Distribution functions near the N terminals $f_{\mathrm{1/2}}(\varepsilon)$ are equilibrium Fermi-Dirac functions with local temperature and chemical potential of the corresponding terminal. EED near S should satisfy Andreev conditions for the energies below the gap:
		\begin{equation}
		f^{*}_{1/2}(\varepsilon) =
		\begin{cases}
		1-f^{*}_{1/2}(-\varepsilon)      & \quad |\varepsilon|<\Delta \\
		(\exp(\varepsilon/{k_{\mathrm{B}}T})+1)^{-1} & \quad |\varepsilon|>\Delta
		\end{cases}
		\label{eq4}
		\end{equation}	
	}
	Following Nagaev and Buttiker \cite{PhysRevB.63.081301} we separately find electron distribution functions for energies below and above the gap and then sew them together.
	{\color{black}With EEDs near S terminal in hand, we are able to calculate current noise spectral density in both NS junction \cite{NAGAEV1992103} i.e. reflected (R) and transmitted (T) shot noise:  
	\begin{equation}
S_{\mathrm{R/T}}=\frac{2}{3}G_{\mathrm{1/2}}\int\Big[2f_{\mathrm{1/2}}(\varepsilon)(1-f_{\mathrm{1/2}}(\varepsilon))+2f_{\mathrm{1/2}}^{*}(\varepsilon)(1-f_{\mathrm{1/2}}^{*}(\varepsilon))+f_{\mathrm{1/2}}(\varepsilon)(1-f_{\mathrm{1/2}}^{*}(\varepsilon))+f_{\mathrm{1/2}}^{*}(\varepsilon)(1-f_{\mathrm{1/2}}(\varepsilon))\Big]d\varepsilon
\label{S5}
\end{equation}	

}

	\subsection{Sub-gap ($|\varepsilon|<\Delta$)}
	\textcolor{black}{First we calculate for energies within the superconducting gap. Since noise temperature gradient is zero at N/S interface, then the correction to the noise temperature due to the finite $r$ is of the second order and we can neglect it. In order to find the energy distributions $f^{*}_{1}(\varepsilon)$ and ${f^{*}_1}(\varepsilon)$ at the two ends of the proximitized wire section, one has to fulfill the continuity of the heat 
		fluxes in these points at any given $\varepsilon$. In this way
		we get the following two equations:
		\begin{equation}
		G_{\mathrm{2}}[f_{\mathrm{2}}(\varepsilon)-f_{\mathrm{2}}^{*}(\varepsilon)]-G_{\mathrm{2}}[f_{\mathrm{2}}(-\varepsilon)-f_{\mathrm{2}}^{*}(-\varepsilon)]=G_{\mathrm{th}}[f_{\mathrm{2}}(\varepsilon)-f_{\mathrm{1}}^{*}(\varepsilon)]-G_{\mathrm{th}}[f_{\mathrm{2}}(-\varepsilon)-f_{\mathrm{1}}^{*}(-\varepsilon)]
		\label{eq5}
		\end{equation}
		\begin{equation}
		G_{\mathrm{1}}[f_{\mathrm{1}}(\varepsilon)-f_{\mathrm{1}}^{*}(\varepsilon)]-G_{\mathrm{1}}[f_{\mathrm{1}}(-\varepsilon)-f_{\mathrm{1}}^{*}(-\varepsilon)]=G_{\mathrm{th}}[f_{\mathrm{2}}(\varepsilon)-f_{\mathrm{1}}^{*}(\varepsilon)]-G_{\mathrm{th}}[f_{\mathrm{2}}(-\varepsilon)-f_{\mathrm{1}}^{*}(-\varepsilon)]
		\label{eq6}
		\end{equation}
		where the terms with $\varepsilon$ correspond to the
		particle heat flux and the terms with $-\varepsilon$ to the hole
		heat flux. The equations~\eqref{eq5} and~\eqref{eq6} are for the right and the left NS interface respectively. For convinience lets introduce  $F(\varepsilon)\equiv f(\varepsilon)- f(-\varepsilon)$, then:}
	\begin{equation}
	\begin{gathered} 
	F^*_1(\varepsilon)=\frac{G_1G_2F_1(\varepsilon)+G_{\mathrm{th}}G_1F_1(\varepsilon)+G_{\mathrm{th}}G_2F_2(\varepsilon)}{G_{\mathrm{th}}G_1+G_{\mathrm{th}}G_2+G_{1}G_{2}},~
	F^*_2(\varepsilon)=\frac{G_1G_2F_2(\varepsilon)+G_{\mathrm{th}}G_2F_2(\varepsilon)+G_{\mathrm{th}}G_1F_1(\varepsilon)}{G_{\mathrm{th}}G_1+G_{\mathrm{th}}G_2+G_{1}G_{2}}
	\end{gathered}
	\end{equation}
	Since $f_{i}^*(\varepsilon)=1-f_{i}^*(-\varepsilon)$, we can easily find $f_i^*=(1+F_i^*)/2$
	\subsection{Above-gap ($|\varepsilon|<\Delta$)}
	\textcolor{black}{To calculate for energies above the superconducting gap, we need to take into account the fact that now there is a finite gradient of noise temperature near N/S interface. Then correction to $T_{\mathrm{N}}$ due to the finite $r$ is of the first order and we can not neglect it. Fortunately, for energies above the gap, S terminal acts as regular normal lead, so we can assume $f^{*}(\varepsilon)=~f^{*}_{1/2}(\varepsilon)$. In order to find $f^{*}(\varepsilon)$, one has to fulfill current conservation law for each energy:
		\begin{equation}
		G_{\mathrm{2}}[f_{\mathrm{2}}(\varepsilon)-f^{*}(\varepsilon)]+G_{\mathrm{1}}[f_{\mathrm{1}}(\varepsilon)-f^{*}(\varepsilon)]=r^{-1}[f^{*}(\varepsilon)-f_{0}(\varepsilon)]
		\end{equation}
		where $f_{0}(\varepsilon)$ is a Fermi-Dirac distribution in in the normal state of grounded Al terminal and $r$ is the interface resistance.}
	\begin{equation}
	\begin{gathered} 
	f^*(\varepsilon)=\frac{G_1f_1(\varepsilon)+G_2f_2(\varepsilon)+1/rf_0(\varepsilon)}{G_1+G_2+r^{-1}}
	\end{gathered}
	\end{equation}
	
	\newpage
{\color{black}	
	\section{\textcolor{black}{Reflected and Transmitted shot noise}}
		\begin{figure*}[h]
		\begin{center}
			\vspace{-0.5cm}
			\includegraphics[scale=1]{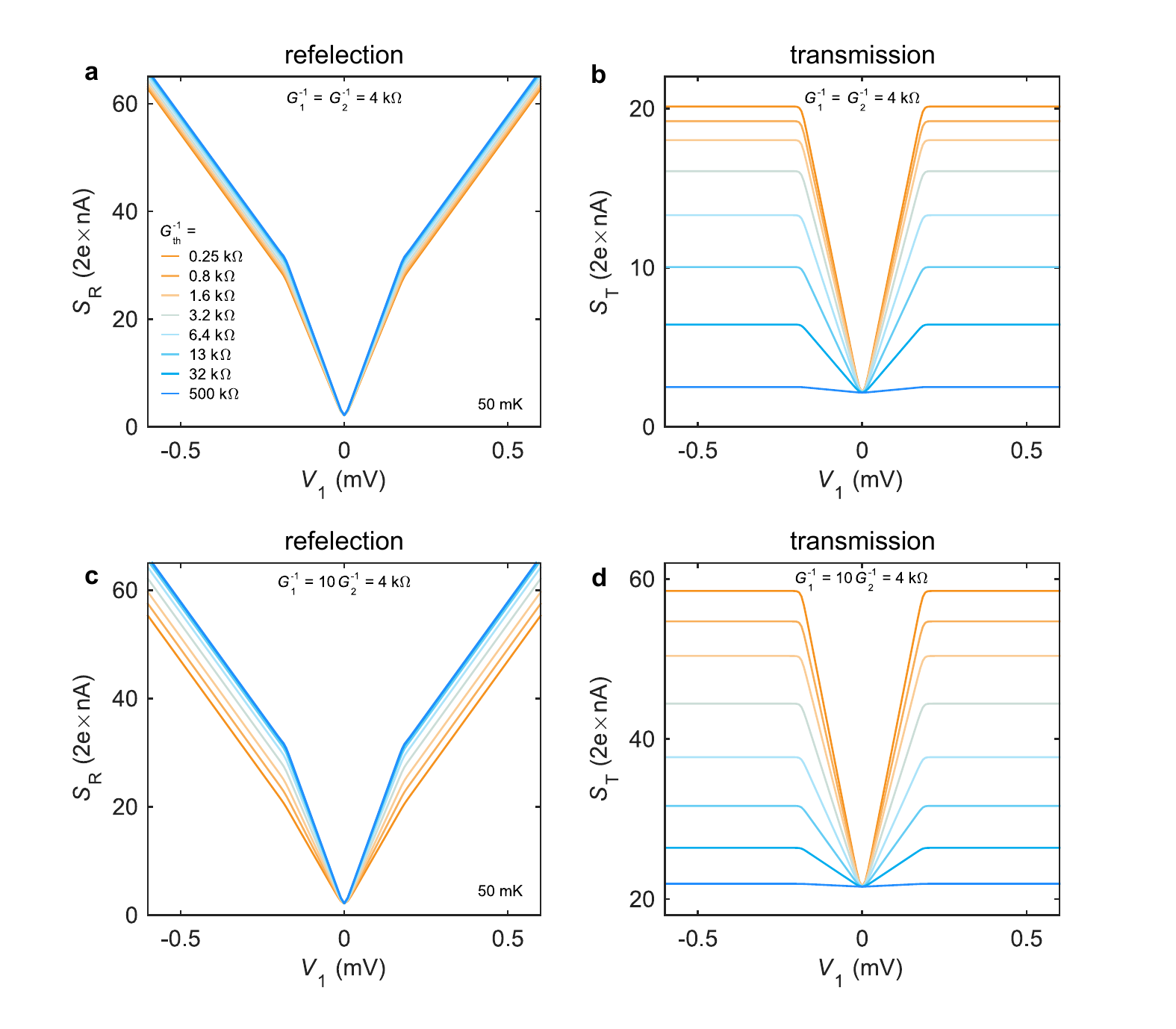}
			\vspace{-0.5cm}
			\caption{\textbf{Analytical model: results.} (a) - (b)~Calculated shot noise in reflection and transmission configurations for $G_{\mathrm{1}}=G_{\mathrm{2}}$ and various $G_{\mathrm{th}}$.~(c) - (d)~Calculated shot noise in reflection and transmission configurations for $G_{\mathrm{2}}=10G_{\mathrm{1}}$ and various $G_{\mathrm{th}}$.}
			\label{fig:T_vs_R}
		\end{center}
	\end{figure*}
Calculated with \eqref{S5} reflected and transmitted shot noise are plotted in fig.\,\ref{fig:T_vs_R}. For clarity, we consider a case of the perfect interface $r=0$ in two limits of symmetric $G_{\mathrm{1}}=G_{\mathrm{2}}$ (a, b) and highly asymmetric $G_{\mathrm{1}}=0.1G_{\mathrm{2}}$ (c,~d) NSN device. Here we vary $G_{\mathrm{th}}$ which is responsible for the heat transmission between two adjacent NS junctions.

For pinched off or extra-long middle section of NSN device $G_{\mathrm{th}}\ll G_{1/2}$, two NS junctions are almost decoupled from each other and the transmitted signal in fig.\,\ref{fig:T_vs_R} b, d  is negligible. At the same time, the reflected noise is following the well-known $e^{*}=2e\to e$ crossover for diffusive NS junctions \cite{PhysRevB.63.081301} that means a complete sub-gap reflection of the heat flux from S terminal. Increasing $G_{\mathrm{th}}$ we allow some heat transmission towards the right N terminal which results in the reduced sub-gap effective charge $e^{*}<2e$ (slope) of the reflected noise. Particularly for the asymmetric device, effective charge approaches a single value $e^{*}\to e$ as $G_{\mathrm{th}},~G_{2}\gg G_{1}$ in fig.\,\ref{fig:T_vs_R}c. This is expected since S terminal in this case is effectively shorted with the right N terminal which serves as a heat sink for sub-gap quasiparticles.

The lost portion of the reflected heat flux is evident from the transmitted signal in fig.\,\ref{fig:T_vs_R}b,~d which increases with increasing $G_{\mathrm{th}}$ at given bias voltage for both symmetric and asymmetric NSN junctions. Being the strong function of thermal conductance, the non-local noise is suitable for accurate determination of $G_{\mathrm{th}}$.
	
}
	\newpage
	
	\section{Noise vs. Resistive thermometry}
	
	\begin{figure*}[h]
		\begin{center}
			\vspace{-0.5cm}
			\includegraphics[scale=1]{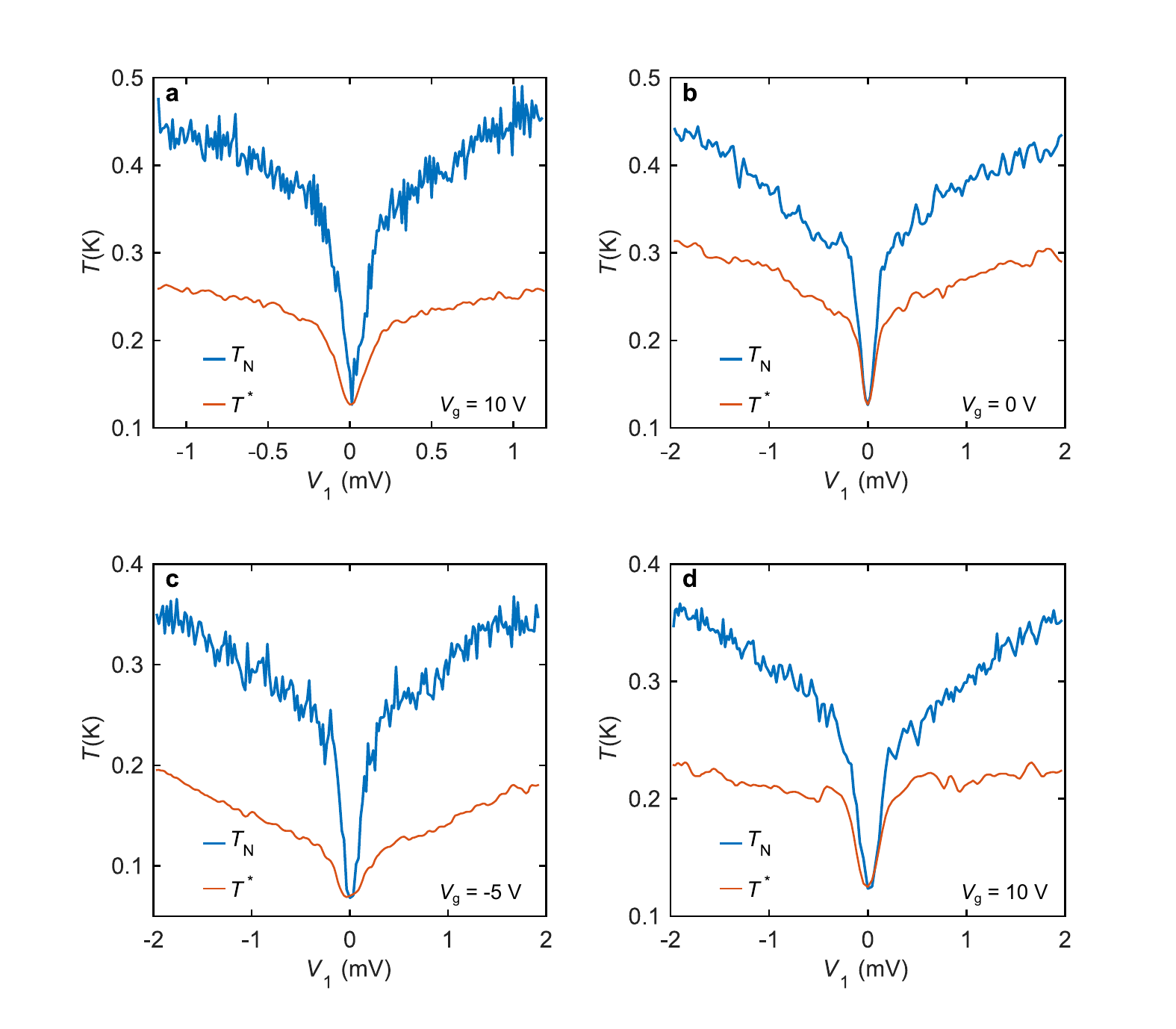}
			\vspace{-0.5cm}
			\caption{\textbf{Comparison of the non-local noise thermometry and resistive thermometry.} (a) - (b)~Measured noise and effective temperature of the floating NS junction (NSN~-~II device) at different back~-~gate voltages. Zero-bias thermometry underestimates noise one by the factor of $\sim2$.}
			\label{fig:t_n_t_star}
		\end{center}
	\end{figure*}
	
	Is this section we highlight the difference between noise thermometry and resistive thermometry approaches.
	
	To describe numerically the resistive thermometry, we consider NS junction as a two-terminal coherent quasi-1D conductor with $n$ channels and energy-dependent diagonal transmission coefficients $Tr_i(\varepsilon)$ connecting two reservoirs with temperatures $T_\mathrm{L}$ and $T_\mathrm{R}$. We now determine the linear response conductance $G_\mathrm{1}$ in such a system, where a temperature bias across the conductor might be present.
	
	If a small voltage bias $dV$ is applied across the conductor, the linear-response current through the nanowire $dI$ is~\cite{BLANTER20001}:
	
	\begin{equation}
	dI = \frac{2e}{h}\sum_{i}^{n} \int Tr_i(\varepsilon) \left[ f\left(\varepsilon-\frac{edV}{2},T_\mathrm{R}\right) - f\left(\varepsilon+\frac{edV}{2},T_\mathrm{L}\right) \right] d\varepsilon,
	\end{equation}
	
	where $f(\varepsilon,T)$ is the equilibrium Fermi-Dirac distribution at temperature $T$. We note, that the conductance temperature dependence in this model is fully enclosed in the energy dependence of $Tr_i(\varepsilon)$. Now we transform this equation by taking out the small bias from the distribution functions and separating the sum:
	
	\begin{equation}
	dI = \frac{2e}{h} \int \left[\sum_{i}^{n} Tr_i(\varepsilon)\right] \left[ \frac{\partial f\left(\varepsilon,T_\mathrm{R}\right)}{\partial \varepsilon}\left(-\frac{edV}{2}\right) - \frac{\partial f\left(\varepsilon,T_\mathrm{L}\right)}{\partial \varepsilon} \left(\frac{edV}{2}\right)\right] d\varepsilon,
	\end{equation}
	
	which leads to the linear response conductance $G_\mathrm{1,non-eq}$ in form:
	
	\begin{equation}
	G_\mathrm{1} = \partial I/\partial V = -\frac{2e^2}{h} \frac{1}{2} \int \left[\sum_{i}^{n} Tr_i(\varepsilon)\right] \left[ \frac{\partial f\left(\varepsilon,T_\mathrm{R}\right)}{\partial \varepsilon} + \frac{\partial f\left(\varepsilon,T_\mathrm{L}\right)}{\partial \varepsilon} \right] d\varepsilon = \frac{G_1(T_\mathrm{R})+ G_1(T_\mathrm{L})}{2}.
	\label{eq:supp_comp3}
	\end{equation}
	
	where $G_1(T)$ is the ordinary equilibrium conductance which can be measured by varying the bath temperature. The (\ref{eq:supp_comp3}) clearly demonstrates, that the temperature, measured via resistive thermometry $T^*$ in the main text obeys a relation:
	
	\begin{equation}
	G_1(T^*) = \frac{G_1(T_\mathrm{R})+ G_1(T_\mathrm{L})}{2},
	\end{equation}
	
	We note, that this result is only valid for the case of equilibrium distribution functions at both nanowire terminals. For the triple-step distribution expected for S terminal in our experiment, the conductance depends on the exact form of $Tr_i(\varepsilon)$.
	
	We now compare this result for $T^*$ with the expected noise temperature $T_\mathrm{N}$ when one terminal has significantly higher temperature ($T_\mathrm{R} \gg T_\mathrm{L}$). For this case $T_\mathrm{N}$ can be expressed in form $T_\mathrm{N} = \alpha T_\mathrm{R}$, with coefficient $\alpha$ depending on the shape of distribution function at the terminal~\cite{Tikhonov2016}. For the double-step distribution $\alpha = 2/3$, while in the case of equilibrium distribution $\alpha=(1+ln(2))/2 \approx 0.56$.
	
	Neglecting the non-linearity of of $G_1(T)$, which is present in experiment (see supplemental fig.~\ref{sup_fig2_new}b), we obtain that the relation $T_\mathrm{N} \approx 4/3 T^*$.
	
	In experiment, however, the discrepancy between $T_\mathrm{N}$ and $T^*$ is more prominent with $T_\mathrm{N}-T_\mathrm{bath} \approx 2 (T^* - T_\mathrm{bath})$ (see supplemental fig.~\ref{fig:t_n_t_star}). Apart from the discussed earlier effect of the non-equilibrium distribution on $G_1$, this inconsistency might be related to the dephasing being present in nanowire. Such dephasing may effectively break the nanowire into several coherent section, with possibly different signs of temperature dependence, leading to a further dampening of $T^*$ compared to $T_\mathrm{N}$.
	
	\newpage
		\section{Non-local voltage generated by temperature bias}
	
	In order to describe the symmetric component of the non-local $I$-$V$s presented in Fig.~4d of the main text we consider a thermoelectric generation of voltage. In the Landauer-B$\rm\ddot{u}$ttiker formalism, the conversion of the temperature bias to electric current is a result of the energy dependence of the eigen-channel transparencies $Tr_i(\varepsilon)$. Thermoelectric current generated in a short-circuited conductor can be written as:
	\begin{equation}
	I_{\mathrm{TE}} = \frac{2e}{h}\sum_{i}^{n} \int Tr_i(\varepsilon) \left[ f\left(\varepsilon,T_\mathrm{R}\right) - f\left(\varepsilon,T_\mathrm{L}\right) \right] d\varepsilon \approx \frac{2e}{h}\sum_{i}^{n} Tr_i^{'} \int \varepsilon\left[ f\left(\varepsilon,T_\mathrm{R}\right) - f\left(\varepsilon,T_\mathrm{L}\right) \right] d\varepsilon, \label{TE1}
	\end{equation}
	where the quasiparticle energy $\varepsilon$ is measured with respect to the chemical potential that is the same for the right and left leads of the conductor. Note that in this equation we approximated the energy dependence $Tr_i(\varepsilon)$ with the lowest order non-vanishing term $Tr_i(\varepsilon)=Tr_i(0)+Tr_i^{'}\varepsilon$, where $Tr_i^{'}\equiv dTr_i(\varepsilon)/d\varepsilon|_{\varepsilon=0}$. It is straightforward to see that eq.~(\ref{TE1}) results in a parabolic T-dependence of $I_{\mathrm{TE}}$:
	\begin{equation}
	I_{\mathrm{TE}} \approx \frac{\pi^2k_B^2}{6}\frac{2e}{h}\sum_{i}^{n} Tr_i^{'} \left[T_\mathrm{R}^2-T_\mathrm{L}^2 \right]. \label{TE2}
	\end{equation}
	Measured thermoelectric voltage that builds up on a floating conductor is simply $V_{\mathrm{TE}}=-I_{\mathrm{TE}}G_0^{-1}$, where ${G_0=2e^2/h\sum_{i}^{n} Tr_i(0)}$ is the linear-response conductance. For a small temperature difference $\Delta T \equiv T_\mathrm{R}-T_\mathrm{L} \ll T$, the eq.~(\ref{TE2}) can be written as:
	\begin{equation}
	V_{\mathrm{TE}} \approx (S/T)T \Delta T, \text{ where } S/T = -\frac{\pi^2k_B^2}{3e} \sum_{i}^{n} Tr_i^{'} \left [\sum_{i}^{n} Tr_i \right]^{-1}.  \label{TE3}
	\end{equation}
	In other words, in this approximation the thermoelectric response is fully characterized by the T-independent Seebeck parameter $S/T$. With this notation we obtain the expression for arbitrary thermal biases on the conductor:
	\begin{equation}
	V_{\mathrm{TE}} \approx (S/T) \frac{T_\mathrm{R}^2-T_\mathrm{L}^2}{2}, \label{TE4}
	\end{equation}
	that is used to fit the data for $V_2^{\mathrm{symm}}$ in Fig.~4d of the main text.
	
		\newpage
		
	\section{\textcolor{black}{Critical Temperature of Al contacts}}
	\begin{figure*}[h]
		\begin{center}
			\vspace{-0.6cm}
			\includegraphics[scale=1]{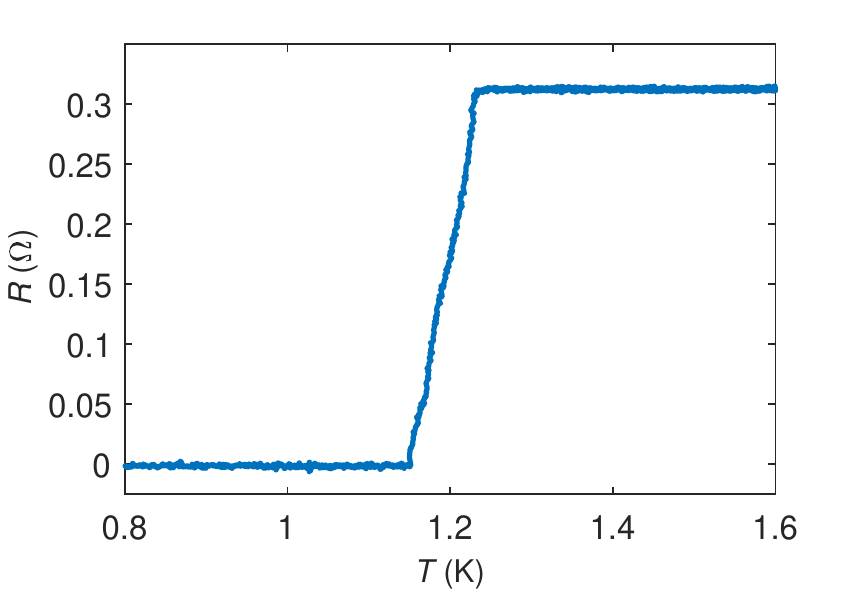}
			\vspace{-0.4cm}
			\caption{\textbf{Superconducting critical temperature of the Al-film.} The resistance of a four-terminal Al strip, deposited via the same process, as the one used in the fabrication of the samples, featured in the main text.}
			\label{sup_fig_al}
		\end{center}
	\end{figure*}
	
	The temperature dependence of superconducting Al, deposited via the same process as described in "Device Fabrication" was performed separately on the four-terminal Al strips, incorporated in the samples studied in~\cite{Bubis_2017}. Here we present raw data (see Fig.~\ref{sup_fig_al}), which leads to the estimate $T_\mathrm{c} = 1.20 \pm 0.03$\,K, based on the position of the middle of transition.

	\section{\textcolor{black}{Device fabrication}}
	{InAs nanowires grown by molecular beam epitaxy on Si substrate~\cite{Hertenberger_2010} are ultrasonicated in isopropyl alcohol. Nanowires are drop casted on Si/SiO2 (300 nm) substrates~\cite{PhysRevB.97.115306} with preliminary defined alignment marks.} For superconducting contacts conventional electron beam lithography (EBL) followed by e-beam deposition of Al (150\,nm) is utilized. {To} obtain {the} ohmic contacts{,} in-situ Ar ion milling is performed before Al deposition in a chamber with a base pressure below $10^{-7}$\,mbar. Normal metal contacts are fabricated in two different ways (different device batches): magnetron sputtering or e-beam deposition. For sputtering (NS and NSN~-~I devices) in-situ Ar plasma etching is followed by sputtering of Ti/Au (5\,nm/200\,nm). Normal metal contacts Ti/Au (5\,nm/150\,nm) in device NSN~-~II are deposited in the same way {as} superconducting ones.
	\renewcommand*{\bibfont}{\small}
	
%